\documentclass[sigconf]{acmart}
\AtBeginDocument{%
  }

\setcopyright{none}
\acmConference[arxiv '26]{arxiv}{2026}{Online}

\usepackage{graphicx}   
\usepackage{hyperref}    
\usepackage{amsmath}
\usepackage{amsfonts}
\usepackage{amsthm}
\usepackage[most]{tcolorbox}
\usepackage{enumitem}
\usepackage{multirow}
\usepackage{booktabs}
\usepackage{tikz}
\usepackage{eso-pic}
\usepackage{xurl}
\usepackage{tikz}
\usetikzlibrary{calc, positioning, arrows.meta, shapes.geometric, fit, backgrounds, shadows}
\usepackage[underline=true,rounded corners=false]{pgf-umlsd}

\definecolor{acmblue}{RGB}{0, 85, 150}
\definecolor{softgray}{RGB}{240, 242, 245}
\definecolor{bordergray}{RGB}{180, 180, 180}
\definecolor{securityred}{RGB}{180, 30, 30}
\definecolor{layerfill}{RGB}{225, 245, 254}

\DeclareMathOperator{\negl}{\textsf{negl}}

\newtcolorbox{mybox}[1][]{colframe=black,colback=gray!5,boxrule=0.6pt,arc=2pt,
                          left=8pt,right=8pt,top=6pt,bottom=6pt,breakable,#1}

\begin{document}

\title{VeriLLM: A Lightweight Framework for Publicly Verifiable Decentralized Inference}

\author{Ke Wang}
\authornote{These authors contributed equally to this work.}
\affiliation{%
  \institution{Gradient}
  \country{United States}
}
\email{wangke@gradient.network}

\author{Zishuo Zhao}
\affiliation{%
  \institution{University of Illinois Urbana-Champaign, Gradient}
  \country{United States}
}
\authornotemark[1]
\email{zishuo2@illinois.edu}

\author{Xinyuan Song}
\affiliation{%
  \institution{Emory University}
  \country{United States}
}
\email{xinyuan.song@emory.edu}

\author{Zelin Li}
\affiliation{%
  \institution{Ohio State University}
  \country{United States}
}
\email{li.15526@osu.edu}

\author{Libin Xia}
\affiliation{%
  \institution{Peking University}
  \country{China}
}
\email{lbxia@stu.pku.edu.cn}

\author{Chris Tong}
\affiliation{%
  \institution{Gradient}
  \country{United States}
}
\email{ChrisT@gradient.network}

\author{Bill Shi}
\authornote{Corresponding author.}
\affiliation{%
  \institution{Gradient}
  \country{Canada}
}
\email{tianyu@gradient.network}

\author{Wenjie Qu}
\authornotemark[2]
\affiliation{%
  \institution{National University of Singapore}
  \country{Singapore}
}
\email{wenjiequ@u.nus.edu}

\author{Eric Yang}
\affiliation{%
  \institution{Gradient}
  \country{United States}
}
\email{eric@gradient.network}

\author{Lynn Ai}
\affiliation{%
  \institution{Gradient}
  \country{United States}
}
\email{lynn@gradient.network}

\renewcommand{\shortauthors}{Ke Wang et al.}

\begin{abstract}
  Decentralized inference provides a scalable and resilient paradigm for serving large language models (LLMs), enabling fragmented global resource utilization and reducing reliance on centralized providers. However, in a permissionless environment without trusted nodes, ensuring the correctness of model outputs remains a core challenge. We introduce \textbf{VeriLLM}, a publicly verifiable protocol for decentralized LLM inference that achieves security 
  with incentive guarantees while maintaining practical efficiency. VeriLLM combines lightweight empirical rerunning with minimal on-chain checks to preclude free-riding, allowing verifiers to validate results at approximately 1\% of the underlying inference cost by exploiting the structural separation between prefill and autoregressive decoding. To prevent verification bottlenecks, we design an isomorphic inference–verification architecture that multiplexes both inference and verification roles across the same GPU workers. This design (i) improves GPU utilization and overall throughput, (ii) enlarges the effective validator set, enhancing robustness and liveness, and (iii) enforces task indistinguishability to prevent node-specific optimizations or selective behavior. Through theoretical analysis and system-level evaluation, we show that VeriLLM achieves reliable public verifiability with minimal overhead, offering a practical foundation for trustworthy and scalable decentralized LLM inference.
\end{abstract}




\maketitle

\section{Introduction}

The recent surge of Large Language Models (LLMs) is driven by their broad utility in diverse domains, including software development~\cite{wang2025enhancingcodellmsreinforcement} and creative content generation, resulting in widespread adoption across both industrial and academic settings~\cite{brown2020language,openai2023gpt4,touvron2023llama}. However, this progress exposes a fundamental weakness in current AI infrastructure: the computational and model resources required to support large-scale inference are heavily centralized within a few dominant technology providers~\cite{bender2021dangers,zaharia2023accelerating,sevilla2022compute}. Such concentration leads to systemic risks, including privacy violations, restricted access, single points of failure, and monopolistic control over core AI capabilities~\cite{bender2021dangers,whittlestone2021sociotechnical,weidinger2023sociotechnical,hazan2023foundation}.

To mitigate these risks, \emph{decentralized inference}~\cite{yuan2022decentralized,morpheus2024,gonka2024,awsneuron2024,mei2024fedmoe,menon2024ravnest,yin2024knowledge,qu2025prompt} has emerged as a promising direction. By distributing computation across independent providers, decentralized systems aim to achieve openness, fault tolerance, and fairer access to AI resources~\cite{allen2022decentralized,kuppuswamy2022decentralizedai,raj2023federated,he2021trustworthy}. This paradigm enables verifiable collaboration among untrusted nodes and democratizes access to model execution, thereby improving transparency and resilience of large-scale AI deployments~\cite{villalobos2022future,bai2023open,kuppuswamy2022decentralizedai,zaharia2023accelerating}.

A central challenge in decentralized inference, however, is \textbf{verifiable correctness}. 
In permissionless environments, providers are economically motivated to skip computation ("lazy workers") or apply hidden optimizations—such as quantization or pruning—that degrade output fidelity~\cite{ho2018security,wang2022publiccheck,scaramuzza2025zkm,ong2025toploc,wang2024modelcompression,xu2023compressprompt}
Without a robust verification mechanism, participants cannot ensure the integrity of model outputs~\cite{castro1999practical,cachin2011introduction,zhang2020delphi,xu2021verifynet,zamani2023decentralized}. 
As a result, independent verification becomes an operational necessity for any decentralized inference infrastructure~\cite{huang2024trustworthiness}.

At a fundamental level, verifiable LLM inference in permissionless environments faces inherent tensions among computational efficiency, numerical fidelity, and economic incentives. Exact verification is infeasible due to floating-point nondeterminism and hardware heterogeneity, while replication-based approaches incur prohibitive cost and collapse under rational laziness (Verifier's Dilemma).
Cryptographic proofs (e.g., ZK-SNARKs) formally attest to correctness but incur overheads several orders of magnitude higher than native inference, rendering them impractical for transformer-scale models~\cite{ezkl-github,chen2024zkml,sun2024zkllm,qu2025zkgpt}.
Consensus-based approaches rely on replication and voting but suffer from critical scalability flaws: they incur multiplicative computational costs, struggle with floating-point non-determinism across heterogeneous devices, and require an impractically large set of honest nodes to maintain security guarantees.
Furthermore, both approaches align poorly with modern blockchain environments, where on-chain verification of large ML proofs is economically infeasible and numerical reproducibility across heterogeneous hardware remains unsolved.

To address these limitations, we introduce \textbf{VeriLLM}, a high-assurance protocol for decentralized inference that achieves publicly verifiability with negligible overhead.
By strategically leveraging the structural asymmetry of LLM execution, VeriLLM reduces the verification burden to approximately 1\% of the inference cost.
Crucially, unlike consensus protocols that demand a synchronous honest majority within every local committee, VeriLLM guarantees security relying only on a relaxed global honest majority assumption. This trust model is realized through a single-round re-verification mechanism.
To further harden this foundation, we design a task-indistinguishable isomorphic network architecture where nodes dynamically multiplex roles. 
This design maximizes the effective validator set and significantly enhances security against selective attacks and resource efficiency.
Furthermore, VeriLLM employs a sampling-based on-chain comparison mechanism that simultaneously resolves the Verifier’s Dilemma (lazy verification) and enables lightweight on-chain settlement, while remaining robust to floating-point non-determinism.

We minimize verification costs by leveraging a structural asymmetry in Transformer execution. 
Since the verifier already possesses the complete output sequence claimed by the worker, it can bypass the expensive, sequential autoregressive decoding process.
 Instead, the verifier concatenates the input prompt with the generated output tokens and processes the entire sequence as a single, fully parallelizable Prefill operation.
 This shifts the workload from a memory-bandwidth-bound serial process to a compute-bound parallel operation that efficiently saturates GPU Tensor Cores.
 Consequently, verification overhead drops to approximately 1\% of the inference cost—an advantage that becomes even more pronounced in long-context tasks, where the marginal cost of parallel verification is negligible.

 To decisively resolve the Verifier’s Dilemma, we enforce a sampling-based audit mechanism. 
 Verifiers are required to submit not only their final judgment but also specific sampled hidden states.
 These lightweight state fragments undergo on-chain adjudication via the smart contract, where any deviation triggers immediate penalization. To ensure comprehensive security, we design a distinct comparison protocol tailored to different adversarial scenarios. 
To address floating-point non-determinism, we introduce a noise-tolerant comparison algorithm specifically optimized for floating-point arithmetic. 
By experimentally calibrating acceptance thresholds based on real-world hardware, this algorithm can effectively filter out numerical divergence, preventing false positives while maintaining strict sensitivity to malicious deviations.

While VeriLLM operates under a global honest-majority assumption, a critical challenge is the risk of 'local majority failure', where a randomized committee happens to be captured by a coordinated adversary. 
To mitigate this, we incorporate a re-verification scheme that allows disputes to be escalated to a newly sampled set of verifiers. A central finding of our work is that a single round of re-verification is mathematically sufficient to secure the protocol. Our game-theoretic analysis confirms that the mere option of this one-time escalation effectively anchors local decisions to the global honest majority; this renders honest voting the unique Nash Equilibrium, guaranteeing robustness without incurring the latency of recursive arbitration.

To guarantee protocol integrity without centralized trust, VeriLLM incorporates a suite of cryptographic primitives. First, we employ Verifiable Random Functions (VRFs) to govern both node selection and hidden-state sampling. This ensures that randomness is publicly verifiable and unpredictable, thereby preventing the scheduler from biasing the audit process. Second, to secure on-chain data submission, we implement a multi-stage commit-reveal scheme. This mechanism forces nodes to bind their results before revealing them, effectively neutralizing 'copycat' attacks (front-running) and preventing ex-post facto modification. Finally, we leverage Merkle Proofs to anchor off-chain data, enabling succinct and gas-efficient on-chain verification of data existence and integrity.

\noindent \textbf{Contributions.} We make the following contributions:
\begin{itemize}[left=0em]


\item \textbf{Minimal Verification Cost:} By exploiting the structural separation of Prefill and Decoder phases in transformer inference, VeriLLM reduces verification overhead to approximately 1\% of full inference cost.

\item \textbf{Global Honest Majority Requirement:} Our design only needs an honest majority among global verification nodes instead of committee nodes, enhancing robustness in verification processes.

\item \textbf{Public and Collusion-and-Lazy-Resistant Verification:} A blockchain-based two-layer verification mechanism ensure that all verifiers perform computations honestly while keeping on-chain operations lightweight. This design also bypasses the \emph{Verifier's Dilemma} via a peer-consistency framework \cite{zhao2024takes} that disincentivizes lazy verification.

\item \textbf{Isomorphic Inference–Verification Network:} The architecture multiplexes inference and verification roles across identical GPU workers, improving throughput and preventing task-specific adversarial behavior.

\item \textbf{Noise-Resistant Design:} Our protocol ensures correctness and soundness even in the presence of machine and floating-point errors.

\item \textbf{Comprehensive Security Evaluation:} We theoretically demonstrate robustness against practical attack vectors and show that VeriLLM provides scalable, low-cost, and publicly verifiable decentralized inference. We implement VeriLLM, and our empirical results substantiate our theoretical claims.
\end{itemize}

\vspace{-1em}
\section{Related Work}
\subsection{Activation Sampling \& Compact Commitment}
Lightweight methods verify inference integrity by selectively committing to critical activation states. TOPLOC/INTELLECT~\cite{ong2025toploc} commits to the top-128 final-layer activations every 32 tokens, compressing data 1000× via polynomial congruence (258 bytes per commitment). However, its verification process requires a trusted third party to re-run the computation and provide a ground truth, which reintroduces a central trust assumption. TensorBlock~\cite{proof-of-cache-github} extends this to KV-cache sampling using deterministic token/layer selection, enabling validators to recompute partial trajectories from checkpoints. VeriSplit~\cite{zhang2024verisplit} supports private outsourcing through linear operator masking: clients add noise to inputs, workers return encrypted activations, and clients locally denoise results using precomputed terms, with Merkle trees enabling partial verification. However, their probabilistic security cannot guarantee the detection of sophisticated model tampering, and input privacy relies on noise injection techniques that are vulnerable to advanced reconstruction attacks.
\vspace{-1em}
\subsection{Zero-Knowledge Machine Learning (ZKML)}
Sumcheck-based Protocols: zkCNN~\cite{liu2021zkcnn} accelerates convolution verification by transforming operations to the frequency domain. VerfCNN~\cite{qu2025verfcnn} achieves theoretical optimal prover performance in convolution verification.  zkLLM~\cite{sun2024zkllm} reduces communication by 90\% for non-arithmetic operations (e.g., Softmax) via the tlookup protocol and exploits translation invariance.
QAP-based Protocols: vCNN~\cite{lee2024vcnn} represents convolutions as Quadratic Polynomial Programs (QPP), reducing multiplicative gates exponentially. pvCNN~\cite{weng2023pvcnn} further optimizes tensor operations via Quadratic Matrix Programs (QMP) and batch verification. ZKML~\cite{chen2024zkml} offers modular circuit design with 43 pre-built layer circuits and automated optimization. zkGPT~\cite{qu2025zkgpt} leverages multi-thread parallelization for LLM inference-proof acceleration. Lookup arguments~\cite{campanelli2024lookup} enable matrix-level privacy with 30\% faster proofs.
VOLE-based Protocols: Mystique~\cite{weng2021mystique} achieves sublinear communication for matrix multiplication using sVOLE-based tensor compression. Recent frameworks~\cite{hao2024scalable} optimize nonlinear functions via table lookups and digital decomposition, reducing constraints by over 99\%.

System \& Hardware Optimization: zkLoRA~\cite{roy2025zklora} supports verifiable fine-tuning via incremental proofs for LoRA modules. EZKL~\cite{ezkl-github} automates ZKP circuit generation from ONNX models. DeepProve~\cite{deep-prove-github} accelerates GPU-based ZKML (54–158× faster proofs via parallelized SNARKs). Sertn~\cite{sertn-avs-github} employs optimistic "Staked Deferred Proofs" with economic slashing for invalid results.
However, these methods impose prohibitive prover costs (hours per inference), require trusted setups, and suffer from rigidity in supporting novel architectures due to circuit specialization.

However, these works come with severe drawbacks, including prohibitive prover costs (often hours per inference), the need for trusted setups in many schemes, and architectural rigidity that hinders support for novel operators due to specialized circuit designs.
\vspace{-1em}
\subsection{Trusted Execution Environments (TEEs)}
Atoma Network~\cite{atoma-infer-github} combines TEE attestation with sampling-based consensus for deterministic execution. evML~\cite{evml-github} uses TEE-based remote attestation (e.g., Google/Apple frameworks) with probabilistic auditing. nesa.ai~\cite{nesa-github} orchestrates heterogeneous TEEs for secure multi-party inference. Phala Network~\cite{phala2024} extends this paradigm by providing a "Confidential AI" cloud platform that enables standard Dockerized AI applications to run within GPU-equipped TEEs. However, they inherit vulnerabilities to side-channel exploits (e.g., cache-timing attacks), centralize trust in proprietary hardware (Intel SGX/ARM TrustZone), and exhibit limited auditability across heterogeneous environments.
\vspace{-1em}
\subsection{Consensus \& Optimistic Approaches}
Verde~\cite{arun2025verde} employs optimistic dispute resolution, recomputing only divergent operators (e.g., single attention heads) to minimize overhead. Mira~\cite{mira-network} leverages multi-model consensus against hallucinations using domain-specific thresholds. Ambient~\cite{ambient-ai} introduces Proof-of-Logits (PoL), hashing intermediate logits for spot-check verification at $<0.1\%$ overhead. However, they introduce delayed finality during dispute windows and implicitly assume honest majority participation.

\vspace{-1em}
\section{Design Goals and Threat Alignment}

This section formalizes the security and system design goals of \textbf{VeriLLM} and establishes their alignment with the assumed adversarial capabilities. 
Our design aims to provide verifiable, tamper-evident, and publicly auditable large language model (LLM) inference in a decentralized environment without relying on trusted majorities or centralized coordination~\cite{cachin2011introduction,allen2022decentralized,zhang2020delphi}.
\vspace{-1em}
\subsection{Adversarial Capabilities}

We consider a permissionless network composed of three types of participants:
\emph{Inferencers}, \emph{Verifiers}, and a \emph{Scheduler}. 
Each participant may deviate from the prescribed protocol to gain computational or economic advantage~\cite{castro1999practical,arun2025verde}.

\begin{itemize}[leftmargin=10pt]
\item \textbf{C1: Malicious Inferencer.} An Inferencer may attempt to reduce computation cost by (i) running a smaller or quantized model, (ii) terminating early, or (iii) reusing cached or fabricated hidden states while pretending to produce valid outputs~\cite{chen2024zkml,ezkl-github}.
\item \textbf{C2: Lazy or Colluding Verifiers.} A Verifier may skip recomputation and submit a verdict opportunistically, or collude with other verifiers or the Inferencer to falsely validate an incorrect computation~\cite{qu2025zkgpt,zamani2023decentralized}.
\item \textbf{C3: Byzantine Scheduler.} The Scheduler may selectively relay, alter, or censor messages; it may attempt to bias role assignment or falsify commitments to conceal misbehavior~\cite{ho2018security}.
\item \textbf{C4: Observation and Timing Adversary.} The adversary can observe network-level messages and timing patterns in an attempt to infer whether a node is serving inference or verification, and may adapt behavior based on such information~\cite{scion2015secure}.
\item \textbf{C5: On-chain Adversary.} Attackers may attempt to bias or replay randomness, forge signatures, or submit inconsistent commitments to gain advantage in commit–reveal process~\cite{micali1999verifiable,bonneau2015sok}.
\end{itemize}

Smart contracts are assumed to execute correctly as specified. 
Cryptographic primitives (hashes, digital signatures, and VRFs) are assumed secure under standard assumptions (collision resistance, EUF-CMA, unpredictability).
\vspace{-1em}
\subsection{Design Goals}

VeriLLM is designed to achieve the following system and security goals in the presence of adversaries with capabilities C1–C5.

\textbf{G1. Public Verifiability.}
\emph{Any participant can verify inference correctness without relying on trust or replication.}
All hidden states and token outputs are bound by cryptographic commitments and, through a commit–then–sample procedure~\cite{sun2024zkllm}, any participant can open selected entries and check them against the model specification. 
This goal directly mitigates C1 and C2 by allowing verifiable detection of computation deviation or lazy verification. 
The verification process is auditable both off-chain (via hidden-state recomputation) and on-chain (via sampled scalar checks).

\textbf{G2. Low Overhead.}
\emph{Verification overhead remains below 1\% of the total inference cost.}
The scheme achieves this by limiting on-chain operations to sampled scalar openings rather than full tensor uploads, and by parallelizing verifier recomputation across layer segments~\cite{rajbhandari2022deep}. 
This goal ensures practical deployability under realistic hardware and network conditions, addressing the performance constraints implied by C1 (economic motivation to cheat).



\textbf{G3. Deterministic Accountability.}
\emph{Every message and hidden-state relay is traceable to a verifiable, signed commitment.}
Each segment boundary state is committed via a Merkle root~\cite{merkle1989certified} and signed by the responsible node. 
These signatures are registered on-chain, ensuring that any alteration, omission, or forgery by a malicious Scheduler (C3) or Inferencer (C1) is cryptographically detectable. 
This binding of state to identity guarantees tamper-evident trace integrity.

\textbf{G4. Task-Type Indistinguishability.}
\emph{Workers cannot distinguish whether a job is an inference or a verification task.}
Inferencers and Verifiers receive identical message formats and API calls from the Scheduler, preventing selective deviation based on task type (mitigating C4). 
A node that attempts to skip verification will be caught with high probability during unpredictable on-chain sampling~\cite{yuan2022decentralized}.

\textbf{G5. Compatibility with Decentralized Deployment.}
\emph{The protocol operates robustly across heterogeneous GPUs and compute environments.}
VeriLLM tolerates bounded floating-point divergence by using calibrated statistical thresholds in hidden-state comparison~\cite{shoeybi2019megatron}. 
This ensures that cross-device verification remains feasible and fair while preventing adversaries from exploiting hardware-level drift to mask malicious changes.
\vspace{-1em}
\subsection{Threat–Goal Alignment}

Table~\ref{tab:threatgoal} summarizes the alignment between adversarial capabilities and the corresponding design goals that mitigate each threat.
\setlength{\abovecaptionskip}{2pt}
\setlength{\belowcaptionskip}{2pt}
\vspace{-0.8em}
\begin{table}[!ht]
  \centering
  \caption{Threat–Goal Alignment Matrix for VeriLLM.}
  \label{tab:threatgoal}
  \resizebox{\columnwidth}{!}{
  \begin{tabular}{l ccccccc}
  \toprule
  \hline
  \textbf{Adversarial Capability} & G1 & G2 & G3 & G4 & G5 & G6 \\
  \midrule
  C1. Malicious Inferencer         & \checkmark &   & \checkmark & \checkmark &   & \checkmark \\
  C2. Lazy/Colluding Verifier      & \checkmark &   & \checkmark &            & \checkmark &            \\
  C3. Byzantine Scheduler          &            &   & \checkmark & \checkmark &            &            \\
  C4. Observation/Timing Adversary &            &   &            &            & \checkmark &            \\
  C5. On-chain Adversary           & \checkmark &   &            & \checkmark &            &            \\
  \hline
  \bottomrule
  \end{tabular}
  }
\vspace{-1em}
\end{table}

The combination of these goals forms a security envelope suited for large-scale decentralized inference. 
Public verifiability (G1) and deterministic accountability (G3) guarantee that every computation step is externally auditable. 
Low-overhead design (G2) ensures practical adoption without compromising latency or throughput. 
The Deterministic Accountability guarantee (G3) provides strong correctness in adversarial environments, while task-type indistinguishability (G4) closes adaptive attack channels. 
Finally, cross-device compatibility (G5) grounds the protocol in realistic heterogeneous hardware deployments~\cite{rajbhandari2022deep,shoeybi2019megatron}. 
Together, these principles ensure that VeriLLM achieves transparent, efficient, and verifiable LLM inference under adversarial conditions typical of open, decentralized systems.

\vspace{-1em}
\section{Background \& Problem Statements}

\textbf{Full-Sequence Prefill for Parallel Verification.} 
Autoregressive large language models (LLMs) rely on causal (left-to-right) self-attention, in which each hidden state $h_t$ depends on all preceding tokens $\{x_1, \dots, x_{t-1}\}$.
This dependency constrains the Decoder to operate sequentially, forming the dominant computational bottleneck in inference. 
Practical serving systems mitigate this through a \emph{Prefill} phase that computes the contextual representations for all prompt tokens in parallel and materializes the key–value (KV) cache, allowing subsequent token-by-token decoding to reuse these cached features~\cite{agrawal2023,shazeer2019fast,rajbhandari2022deep}.
At verification time, this architectural separation becomes even more advantageous: since the verifier observes the full input–output sequence $(x_1, \dots, x_T)$, it can execute a single forward pass under the causal mask to recover all hidden states simultaneously, without performing the autoregressive loop.

Formally, let $\mathcal{F}_{\theta}$ denote the LLM parameterized by $\theta$, and $H = \mathcal{F}_{\theta}(X)$ represent the sequence of hidden activations for token matrix $X \in \mathbb{R}^{T \times d_{\text{model}}}$ under causal masking. 
Given the full sequence, the verifier computes:
\begin{equation}
H' = \mathcal{F}_{\theta}(X_{\text{prompt}}   \Vert   X_{\text{output}}),
\end{equation}
where $\Vert$ denotes concatenation. 
Since $H'$ deterministically reconstructs all token representations, it serves as a sufficient statistic for verifying computational consistency. 
Let $C_{\text{verify}}$ and $C_{\text{infer}}$ denote the respective computational costs of verification and inference. 
Empirically, the ratio satisfies
\begin{equation}
\frac{C_{\text{verify}}}{C_{\text{infer}}} = \frac{T_{\text{prompt}}}{T_{\text{prompt}} + T_{\text{output}}} = 0.01,
\end{equation}
indicating that verification requires only about $1\%$ of total inference compute. 
This structural reuse renders verification nearly cost-free while preserving full fidelity of hidden-state comparison.

\textbf{Lazy Verification.}
A naive verification baseline re-executes the entire model forward pass under identical hyperparameters and compares the resulting hidden states with those reported by the Inferencer. 
However, because honest inference is the common case, verifiers have an incentive to skip computation and submit a trivial \texttt{True} verdict—an instance of the \emph{lazy-verification problem}, aka. the \emph{verifier's dilemma} \cite{luu2015demystifying,zhao2024takes}.
Such behavior gradually undermines system integrity and leads to unverifiable results.

To address this, VeriLLM employs a lightweight \emph{commit–sample–check} protocol. 
Each verifier $v_i$ commits to its hidden states by posting a Merkle root $r_i = \mathsf{Merkle}(H_i)$~\cite{merkle1989certified}. 
After commitments are finalized, a smart contract derives random challenge indices $\mathcal{S} = \{s_1, \dots, s_k\}$ from a verifiable random function (VRF)~\cite{micali1999verifiable}, ensuring unpredictability and fairness. 
Each verifier must open the corresponding leaves $\{H_i[s_j]\}_{j=1}^k$ and submit inclusion proofs $\pi_i[s_j]$ to the contract. 
Let $\delta(H_i[s_j], H^*[s_j])$ denote the numerical deviation from the Inferencer’s committed states. 
The contract enforces:
\begin{equation}
\mathsf{Slash}(v_i) =
\begin{cases}
1, & \text{if } \exists j \text{ s.t. } \delta(H_i[s_j], H^*[s_j]) > \epsilon,\\
0, & \text{otherwise,}
\end{cases}
\end{equation}
where $\epsilon$ is a tolerance threshold capturing floating-point nondeterminism~\cite{dettmers2022llmint8,frantar2023gptq}. 
Verifiers whose openings align with the Inferencer’s states within $\epsilon$ receive rewards, while inconsistent openings trigger slashing. 
To ensure continuous participation, a peer-consistency mechanism statistically compares verifier outputs and rewards correlation with the empirical distribution expected from honest computation, thus discouraging collusion and inactive nodes.

\textbf{Security under Global Honest Majority.} 
Conventional consensus protocols~\cite{castro1999practical,cachin2011introduction} rely on an honest-majority assumption among sampled nodes to guarantee safety and liveness. 
In contrast, VeriLLM achieves correctness under a more realistic \emph{global honest majority} assumption. 
Let $n$ denote the total number of verifiers, $f$ the number of malicious ones, $k$ the committee size for sampling-based verification, and we assume $k$ is odd. Then in conventional protocols, the security fails if and only if more than $\frac{k}{2}$ committee nodes are malicious. Hence, we have:
\begin{equation}
\tilde{P}_{\text{fail}} = \frac{\sum_{t=\lceil k/2\rceil}^{k}\binom{f}{t}\binom{n-f}{k-t}}{\binom{n}{k}},
\end{equation}
which is the probability that malicious nodes take up the majority of sampled nodes. While this probability is non-trivial even if $f<\frac{n}{2}$, our design can ensure security via the \emph{disputing} design, with the intuition that as long as the honest nodes take up the majority of global nodes, the re-verification will be more favorable to honest inferencers than dishonest inferencers, so we can design proper incentives such that inferencers will be incentivized to dispute \emph{if and only if} they are honest but rejected. The detailed analysis is deferred to Appendix~\ref{app:game}.

Thus, VeriLLM maintains strong probabilistic safety guarantees without global replication. 
To further reduce risk under targeted committee attacks, the system employs a dynamic verifier-selection mechanism that re-samples verifiers over epochs, ensuring that adversarial control of a fixed subset cannot persist across rounds. 
Together with a re-verification fallback procedure, this mechanism ensures to detect and reports inconsistencies with overwhelming probability.

\vspace{-0.5em}
\section{Protocol Design}
\begin{figure}[!ht]
  \centering
  \includegraphics[width=.5\textwidth]{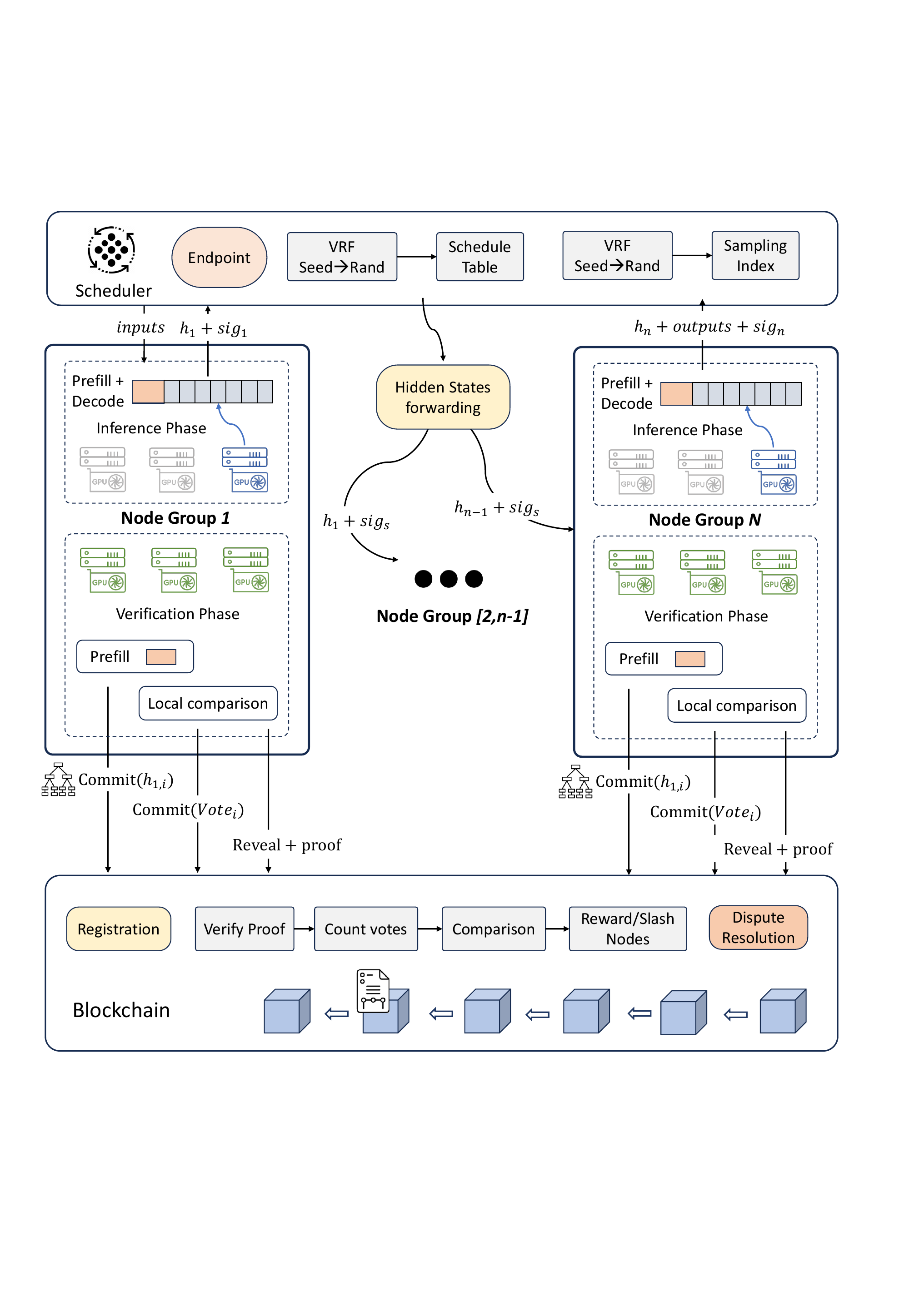}
  \caption{System architecture of VeriLLM. 
  The scheduler employs a verifiable random function (VRF) for unbiased node selection and hidden-state sampling. 
  Each node group performs prefill and decoding, commits hidden states, and submits proofs to the blockchain, 
  which verifies results, counts votes, and applies reward or penalty mechanisms to ensure verifiable decentralized inference.}
  \label{fig:arc}
    \Description{}
    \vspace{-2em}
\end{figure}
\vspace{-0.5em}
\subsection{System Overview}

VeriLLM is a publicly verifiable, decentralized inference framework for large language models (LLMs) built atop a blockchain substrate. The system targets three objectives: publicly auditable correctness, low verification overhead compatible with real-time inference, and transparent incentive enforcement via smart contracts. To achieve these goals, VeriLLM consists of three core components: a \emph{scheduler}, a set of \emph{homogeneous node groups}, and \emph{on-chain verification contracts}, which jointly coordinate inference execution, verification, and rewards to ensure correctness and accountability.

\begin{itemize}[left = 0em]
    \item \textbf{Scheduler.}
    The scheduler orchestrates task placement, resource allocation, and verification planning. For each inference job, it selects candidate node groups and assigns one \emph{Inferencer} and multiple \emph{Verifiers} per group. Although logically centralized, its role assignment and checkpoint sampling are publicly verifiable, as they are derived from a verifiable random function (VRF)~\cite{micali1999verifiable}. This design prevents adversaries from predicting verifier roles or biasing selection. The scheduler also commits job metadata and Merkle roots of intermediate states to the blockchain for immutable traceability.

    \item \textbf{Node Groups.}
    Computation is organized into groups of homogeneous GPU workers, each corresponding to identical Transformer layers and execution configurations, consistent with modern tensor- and pipeline-parallel inference~\cite{rajbhandari2022deep,shoeybi2019megatron}. Nodes may act as either \emph{Inferencers} or \emph{Verifiers} depending on the scheduler’s assignment. To mitigate role-based adversarial behavior, both roles share an identical execution interface, making inference and verification indistinguishable to the worker during the Prefill phase. This \emph{task-type indistinguishability} discourages strategic deviation, as misbehavior risks immediate detection.

    \item \textbf{Verification Contracts.}
    Verification is enforced through on-chain smart contracts~\cite{buterin2014next,wood2014ethereum}. Nodes stake collateral upon participation, and contracts automatically distribute rewards or penalties based on verification outcomes. To reduce gas costs, VeriLLM adopts a two-tier design: computation and Merkle proof generation occur off-chain, while only compact digests, VRF proofs, and sampled Merkle openings are submitted on-chain. This ensures full public verifiability with on-chain costs sublinear in model size and enables trustless third-party auditing via contract event logs.
\end{itemize}

Overall, VeriLLM integrates cryptographic randomness, hardware symmetry, and economic incentives within a unified architecture. The scheduler enforces non-manipulable randomness, node groups provide execution indistinguishability, and smart contracts ensure immutable verification and incentive enforcement. Together, these components enable decentralized inference to achieve correctness and efficiency comparable to centralized systems while remaining fully open and publicly verifiable.

\textbf{End-to-end workflow.}
(1) A client submits an inference request, after which the scheduler selects node groups and assigns Inferencers and Verifiers using a VRF.
(2) The Inferencer executes the forward pass and streams committed hidden states to the scheduler.
(3) Verifiers independently reproduce and compare the corresponding hidden states and submit results on-chain.
(4) The scheduler samples hidden-state indices to prevent lazy verification, and the smart contract finalizes rewards or penalties based on the on-chain comparison.
\vspace{-1em}
\subsection{Registration}

VeriLLM employs an on-chain registration mechanism to ensure that all participating entities---both schedulers and worker nodes---are cryptographically identifiable and economically accountable. Registration is enforced through a family of smart contracts that manage collateral, metadata, and state transitions. The process provides two essential guarantees: (i) each participant is uniquely bound to a verifiable cryptographic identity, and (ii) any detected misbehavior can be penalized by slashing its locked stake.

\textbf{Scheduler registration.}
Schedulers must register on-chain before they are permitted to coordinate inference tasks. Each registration transaction includes: 
\begin{enumerate}
    \item the scheduler’s VRF verification public key, enabling third parties to validate the pseudorandom outputs used in node assignment and hidden-state sampling~\cite{micali1999verifiable};
    \item a network endpoint or relay address to facilitate authenticated communication with worker nodes; and
    \item an initial collateral deposit, ensuring financial accountability for role mismanagement or assignment manipulation.
\end{enumerate}
The VRF key ensures that all sampling and scheduling randomness remains publicly auditable and resistant to precomputation or selective bias~\cite{buterin2014next,wood2014ethereum}.  

\textbf{Node registration.}
Each worker node must likewise register its model configuration and communication endpoint. The registration payload includes:
\begin{enumerate}
    \item a descriptor of the hosted model $m$, including architecture identifier and parameter hash, and
    \item the specific slice $\zeta$ of the model it serves (e.g., Transformer layer range, tensor block, or mixture-of-experts shard).
\end{enumerate}
Nodes also deposit a collateral stake that may be forfeited upon evidence of incorrect inference, falsified verification results, or collusion.  
To ensure fairness and liveness, the registration contract tracks time-stamped updates for all node metadata, allowing the scheduler to continuously maintain an up-to-date view of system membership.

\textbf{Withdrawal and penalties.}
Registered entities may deregister and reclaim their stake, but withdrawals are subject to a fixed unbonding period $\Delta_{\text{wait}}$.  
This grace interval ensures that any pending disputes or verification results can still trigger penalties before funds are released.  
Let $t_{\text{dereg}}$ denote the deregistration timestamp. A withdrawal is finalized only after $t_{\text{final}} = t_{\text{dereg}} + \Delta_{\text{wait}}$, ensuring causal closure of all ongoing verifications.  

This mechanism parallels the delayed-withdrawal model of accountable distributed systems~\cite{castro1999practical,luu2016smartpool}, guaranteeing that both schedulers and workers remain economically bound to their past behavior.
\vspace{-1em}
\subsection{Scheduling Scheme}

The Scheduler coordinates node selection and hidden-state (HS) forwarding across a pipeline of model slices, ensuring efficient inference and verifiable parallelism.  
While logically centralized for low-latency operation, all scheduling decisions are derived from verifiable randomness to prevent manipulation or bias.  

\textbf{Verifiable randomness and auditability.}
Each scheduling decision is deterministically derived from a publicly verifiable random beacon instantiated via a verifiable random function (VRF)~\cite{micali1999verifiable}.  
The Scheduler’s VRF verification key is registered on-chain, and every scheduling round includes the tuple: $(\texttt{job\_id}, \texttt{VRF\_out}, \texttt{VRF\_proof})$, where $\texttt{VRF\_out}$ encodes the pseudorandom seed used for node assignment and hidden-state sampling, and $\texttt{VRF\_proof}$ allows any observer or contract to independently validate correctness.  
This construction guarantees that role assignment and verification checkpoints are unpredictable yet verifiable, preventing any party from biasing job allocation.

\textbf{Node state maintenance.}
The Scheduler maintains a continuously updated view of the network’s active nodes by monitoring on-chain registration events.  
Let $\mathcal{N}$ denote the global set of registered nodes.  
Each node $n \in \mathcal{N}$ exposes metadata fields $\texttt{model}(n)$ and $\texttt{slice}(n)$, describing its hosted model and layer range, respectively.  
The Scheduler constructs homogeneous groups as:
\begin{equation}
G_{m,\zeta} = \{  n \in \mathcal{N}  |  \texttt{model}(n)=m,~\texttt{slice}(n)=\zeta  \},
\end{equation}
where each group represents a cohort of workers hosting identical model slices.  
For a model $m$ with $L$ slices, the pipeline composition is:
\begin{equation}
\mathcal{P}(m) = \langle G_{m,1}, G_{m,2}, \dots, G_{m,L} \rangle.
\end{equation}
When serving a new inference request for model $m$, the Scheduler randomly selects one \emph{Inferencer} and $k$ \emph{Verifiers} within each group $G_{m,\zeta}$ using the VRF seed.  
The assignments are thus both deterministic (given the VRF output) and unpredictable (prior to seed revelation).  

\textbf{Latency and flow control.}
The Scheduler coordinates hidden-state (HS) forwarding between consecutive groups.  
Let $T_\zeta$ denote the transmission latency between $G_{m,\zeta}$ and $G_{m,\zeta+1}$.  
The expected end-to-end latency $T_{\text{total}}$ for an $L$-slice pipeline is:
\begin{equation}
T_{\text{total}} = \sum_{\zeta=1}^{L} \big(T_\zeta + T_{\text{compute}}^{(\zeta)}\big),
\end{equation}
where $T_{\text{compute}}^{(\zeta)}$ is the average inference time per slice.  
Because role assignment is randomized per job, load balancing is implicitly achieved, reducing variance in $T_{\text{compute}}^{(\zeta)}$.  
This property ensures that verification and inference remain tightly coupled without imposing additional synchronization barriers.

In summary, the Scheduler functions as a lightweight but verifiably accountable control plane: it ensures transparent randomness, deterministic scheduling, and reproducible verifier selection, all while preserving the performance characteristics of centralized inference.

\vspace{-1em}
\subsection{Task Scheduling and Hidden-State Management}

\textbf{Task Scheduling.}
Upon receiving a user request $\texttt{req}$, the Scheduler identifies a sequence of node groups 
$\langle G_{m,1}, G_{m,2}, \dots, G_{m,L}\rangle$ that collectively host the model segments required to execute the target model $m$. 
Each group $G_i$ comprises $L_i = |G_i|$ homogeneous nodes, each storing an identical slice of the model parameters and runtime configuration.
To ensure that role assignment is unpredictable yet publicly auditable, the Scheduler derives a per-stage random seed using a verifiable random function (VRF) bound to the request hash, stage index, and group cardinality:
\begin{equation}
h \gets H(\texttt{req}), \qquad 
r_i \gets \mathrm{VRF}(sk_{\mathsf{sch}},   h,   i,   L_i),
\end{equation}
where $sk_{\mathsf{sch}}$ is the Scheduler's VRF secret key, and the corresponding verification key $vk_{\mathsf{sch}}$ is recorded on-chain. 
The Scheduler publishes $(r_i, \pi_i)$, where $\pi_i$ is the VRF proof that allows any observer to validate:
\begin{equation}
\mathrm{VRF.Verify}(vk_{\mathsf{sch}},   \langle h,i,L_i\rangle,   r_i,   \pi_i) = 1.
\end{equation}
The seed $r_i$ initializes a cryptographically secure pseudorandom number generator (CSPRNG), such as a Fisher–Yates permutation seeded with $\mathrm{HKDF}(r_i)$, producing an unbiasable permutation $\sigma_i$ over the node indices in $G_i$. 
From this permutation, the Scheduler deterministically assigns one \emph{Inferencer} and $k$ \emph{Verifiers}:
\begin{equation}
n^{\mathrm{inf}}_i = G_{m,i}[\sigma_i(1)], 
\qquad
\{n^{\mathrm{ver}}_{i,1}, \dots, n^{\mathrm{ver}}_{i,k}\} = G_{m,i}[\sigma_i(2..k{+}1)].
\end{equation}
Because each seed $r_i$ is cryptographically bound to $(h,i,L_i)$ and its proof is anchored on-chain, the resulting assignments are verifiable, reproducible, and non-manipulable. 
Any deviation by the Scheduler or a colluding node is thus immediately detectable through public audit.

\textbf{Hidden-State Forwarding.}
Once role assignment is finalized, the Scheduler coordinates distributed inference as a pipelined relay across the $L$ model segments.  
Let $t$ denote the current decoding step. 
At each step, the pipeline executes as follows:
\begin{itemize}[left = 0em]
    \item The Scheduler transmits the active token sequence $\mathbf{y}_t$ (with $\mathbf{y}_1$ as the initial prompt) to the first segment’s Inferencer ($i=1$). 
    This node computes its hosted layers to produce the hidden-state tensor $S^{(t)}_{2}$.
    \item The Scheduler forwards $S^{(t)}_{2}$ to the next Inferencer ($i=2$), which computes $S^{(t)}_{3}$, and so on through the pipeline.
    \item The relay proceeds until the final Inferencer ($i=L$) produces the terminal hidden state $S^{(t)}_{L+1}$, evaluates the output logits, and predicts the next token $y_{t+1}$, which is returned to the Scheduler.
    \item The Scheduler delivers $y_{t+1}$ to the user and re-injects it as input for the next iteration ($t \leftarrow t+1$).
\end{itemize}
The loop terminates when $y_t = \langle \mathrm{EOS} \rangle$. 
This hop-by-hop relay architecture enables fine-grained auditing and fault isolation while maintaining end-to-end latency comparable to centralized inference. 
Each intermediate tensor $S^{(t)}_i$ is checkpointed for potential verification, ensuring deterministic reproducibility without interrupting the decoding stream.

During each decoding iteration $t$, the overall pipeline computation proceeds as follows.
The Scheduler receives the current token sequence $\mathbf{y}_t$ (where $t{=}1$ corresponds to the prompt) and forwards it to the first segment’s Inferencer.
The first segment embeds and processes $\mathbf{y}_{<t}$ to produce the intermediate hidden state:
\begin{equation}
S^{(t)}_{2} = g_1\big(\mathrm{Embed}(\mathbf{y}_{<t})\big).
\end{equation}
Each subsequent segment $i=2,\dots,L$ receives the preceding output $S^{(t)}_{i}$, executes its corresponding model layers, and produces: $S^{(t)}_{i+1} = g_i(S^{(t)}_{i})$. At the final stage, the tail segment computes the logits and predicts the next token:
\begin{equation}
y_{t+1} = \arg\max f_{\mathrm{LM}}\big(S^{(t)}_{L+1}\big).
\end{equation}
The newly generated token $y_{t+1}$ is then appended to the existing sequence, forming $\mathbf{y}_{\le t} = \mathbf{y}_{<t} \Vert y_t$, which is passed back to the first segment to initiate the next decoding step. 
If $y_t = \langle \mathrm{EOS} \rangle$, the inference process halts.

\textbf{Hidden-State Logging and Commitment.}
To enable verifiable post-hoc auditing, each Inferencer logs its intermediate hidden states for every token and commits to them using cryptographic proofs.  
For each token index $t$, an Inferencer serializes its output tensor $S^{(t)}_i \in \mathbb{R}^{d_k \times d_v}$, constructs a Merkle tree $\mathsf{Merkle}(S^{(t)}_i)$ over all scalar entries, and submits the Merkle root $\gamma^{(t)}_i$ accompanied by a bound signature: $\texttt{Sig}_{\mathrm{inf}}\big(h, i, t, \gamma^{(t)}_i\big)$. To prevent tampering or substitution during relay, the Scheduler also signs the forwarded root: $\texttt{Sig}_{\mathrm{sch}}\big(h, i, t, \gamma^{(t)}_i\big)$, ensuring accountability for both the producer and the forwarder.
This dual-signature commitment model provides end-to-end traceability: any discrepancy between a segment’s output and the forwarded record can be attributed with cryptographic certainty.  
Because verifiers only require Merkle openings for sampled positions, the bandwidth and on-chain verification costs scale sublinearly with tensor size.  
The integration of deterministic serialization, Merkle-based commitment~\cite{merkle1989certified}, and chained signatures provides tamper-evident provenance for all hidden states without introducing measurable latency into the inference path.

Overall, this architecture ensures that every stage of distributed inference remains cryptographically auditable while preserving throughput and latency comparable to optimized large-model serving systems~\cite{rajbhandari2022deep,shoeybi2019megatron}.
\vspace{-1em}
\subsection{Inference Scheme}

\begin{figure}[!ht]
  \centering
  \includegraphics[width=.45\textwidth]{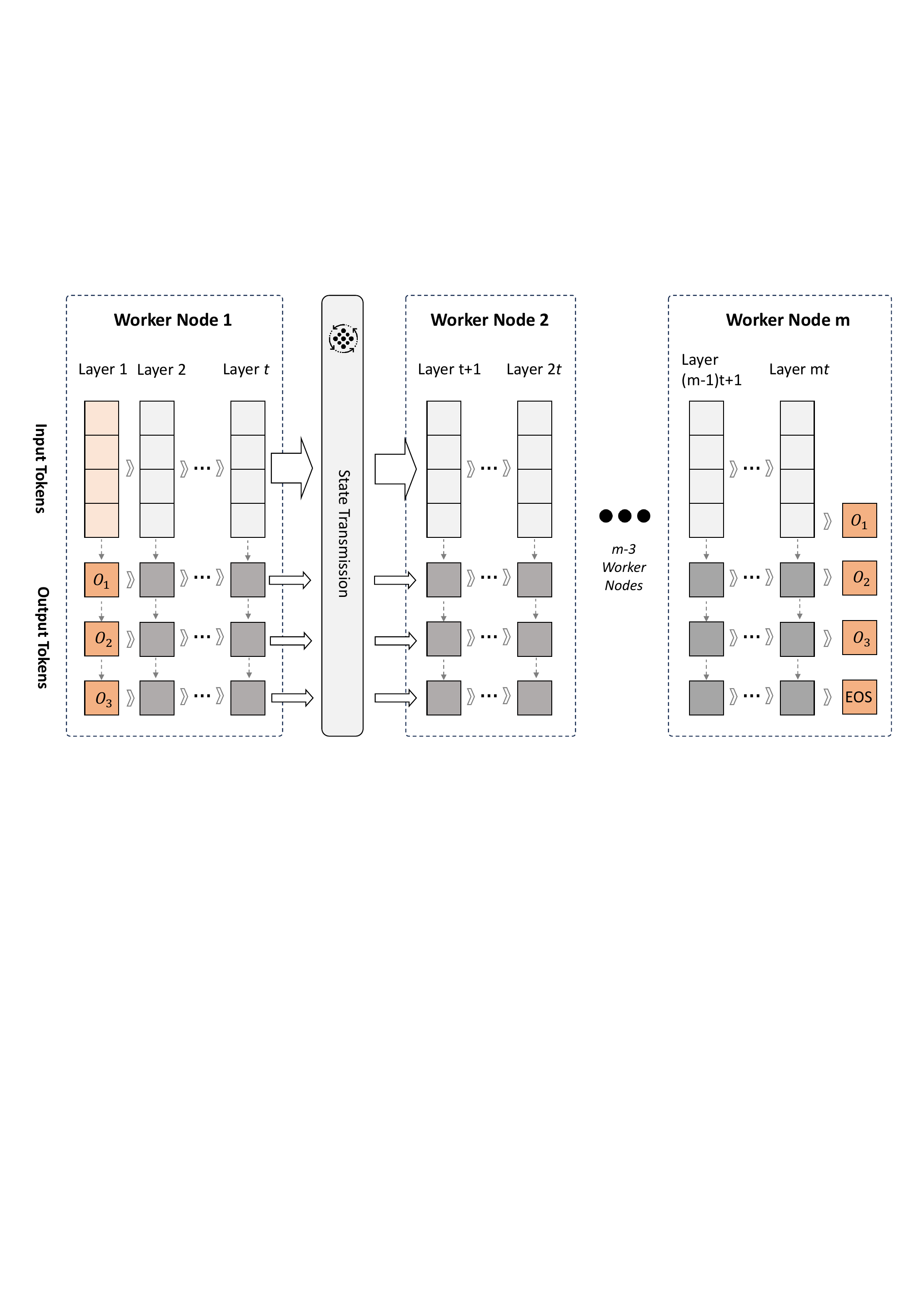}
  \caption{Overview of the decentralized inference architecture. Each segment hosts identical Transformer layers, and hidden states are relayed across the pipeline.}
  \label{fig:arc2}
  \Description{}
  \vspace{-2em}
\end{figure}

The inference process executes online, autoregressive generation over a decentralized, pipelined fabric of homogeneous node groups. 
Each group hosts an identical copy of a contiguous Transformer block, including weights, normalization statistics, and kernel configurations~\cite{shoeybi2019megatron,rajbhandari2022deep}. 
For each incoming task, the Scheduler selects eligible groups and, through a verifiable random function (VRF)~\cite{micali1999verifiable}, assigns one \emph{Inferencer} per group to execute the live forward pass. 
Verifier roles are simultaneously drawn from the same VRF distribution but remain inactive during this phase. 
Throughout generation, the Scheduler relays and logs hidden states across segment boundaries, appending cryptographic commitments and signatures to construct an auditable trace that serves as the foundation for subsequent verification~\cite{zhang2020delphi,chen2024zkml}.

Formally, let the model be divided into $L$ contiguous segments.
At decoding step $t$, given the causal input $x_{1:t} = [\texttt{prompt}   \|   y_{1:t-1}]$, each segment computes its portion of the forward pass as follows:
\begin{align}
S^{(t)}_1 &\leftarrow \texttt{Embed}(x_{1:t}),\\
S^{(t)}_{i+1} &\leftarrow F_i\big(S^{(t)}_i; W_i\big), \quad i = 1, \dots, L,\\
y_t &\leftarrow \texttt{Decode}\big(S^{(t)}_{L+1}\big),
\end{align}
where $S^{(t)}_i$ is the hidden state entering segment $i$, $F_i$ denotes the segment’s forward operator, $W_i$ its parameters, and $\texttt{Decode}$ the final output head combined with the sampling rule. 
The resulting pipeline achieves token-serial yet segment-parallel execution, synchronized by the Scheduler’s relay of hidden states~\cite{rajbhandari2022deep}.

\textbf{Scheduler-Driven Role Assignment.}
For each inference task, the Scheduler enumerates all node groups capable of collectively serving the model and assigns roles in a publicly verifiable manner. 
Using a VRF bound to the task identifier~\cite{micali1999verifiable}, it deterministically selects one Inferencer per group, while keeping the randomness unpredictable before the VRF proof is revealed.
This mechanism ensures that role assignment cannot be biased or precomputed and that every selection remains verifiable through the VRF proof posted on-chain~\cite{bonneau2015sok,qu2025zkgpt}.

\textbf{Pipelined Forward and State Relay.}
During decoding step $t$, segment $i$’s Inferencer receives from the Scheduler both the upstream hidden state $S^{(t)}_i$ and a tamper-evidence package containing a commitment to $S^{(t)}_i$ and a proof of authenticity from the previous node. 
The Inferencer then:
\begin{itemize}[left = 0em]
    \item performs the forward pass to produce $S^{(t)}_{i+1}$;
    \item computes a cryptographic commitment to $S^{(t)}_{i+1}$ (e.g., a Merkle root over a deterministic serialization~\cite{merkle1989certified}) and signs the root with its node key; and
    \item returns the tuple $\{S^{(t)}_{i+1}, \texttt{root}^{(t)}_i, \sigma^{(t)}_i\}$ to the Scheduler.
\end{itemize}
The Scheduler verifies each signature against the on-chain registry (optionally through a zero-knowledge membership proof~\cite{chen2024zkml,sun2024zkllm} to preserve role anonymity), appends its own relay signature, persists the state to stable storage, and forwards the package to the next segment’s Inferencer.
To minimize latency, each receiving node can begin its computation immediately while performing signature verification asynchronously; if any proof later fails, the Scheduler aborts the affected pipeline and invalidates dependent results.

\textbf{Commitment and Signature Discipline.}
At every token step and segment boundary, the produced hidden state is cryptographically anchored by:
\begin{enumerate}
    \item a commitment $\texttt{root}^{(t)}_i = \texttt{MerkleRoot}(S^{(t)}_{i+1})$, and
    \item a node signature $\sigma^{(t)}_i = \texttt{Sig}(\texttt{sk}_i,   \texttt{root}^{(t)}_i)$.
\end{enumerate}
All node public keys are registered under an on-chain Merkle accumulator, enabling compact membership proofs~\cite{micali1999verifiable,bonneau2015sok}. 
Instead of forwarding raw signatures, the Scheduler may alternatively issue a zero-knowledge proof attesting that it holds a valid signature from some registered key on the advertised root—thereby maintaining verifiability without revealing node identity~\cite{sun2024zkllm,qu2025zkgpt}.
Each forwarded artifact is re-signed by the Scheduler to guarantee origin authenticity and replay protection. 
This layered commitment and signature discipline ensures that every hop in the relay chain is both tamper-evident and cryptographically attributable.

\textbf{Autoregressive Token Emission and Feedback.}
The terminal segment ($i=L$) outputs logits or the final hidden state for token $t$, from which the next token $y_t$ is sampled according to the decoding policy~\cite{brown2020language}. 
The Scheduler collects $y_t$ and feeds it back into the pipeline to form $x_{1:t+1}$, advancing the process to step $t{+}1$. 
The loop continues until the model emits an end-of-sequence token $\langle \texttt{EOS} \rangle$ or another termination condition is met.

\textbf{Task-Type Indistinguishability and Robustness.}
The communication protocol, message format, and API surface used by Inferencers during inference are identical to those used by Verifiers in the subsequent verification phase~\cite{yuan2022decentralized}. 
Consequently, no node can distinguish in real time whether it is performing live inference or participating in a verification probe.
This indistinguishability eliminates the opportunity for strategic deviation and simplifies implementation symmetry across both phases.
Combined with VRF-driven role assignment, signed commitments, and auditable relays, the inference scheme produces a verifiable, tamper-evident execution trace with negligible online overhead.

\subsection{Verification Scheme}

\begin{figure}[!ht]
  \centering
  \includegraphics[width=.4\textwidth]{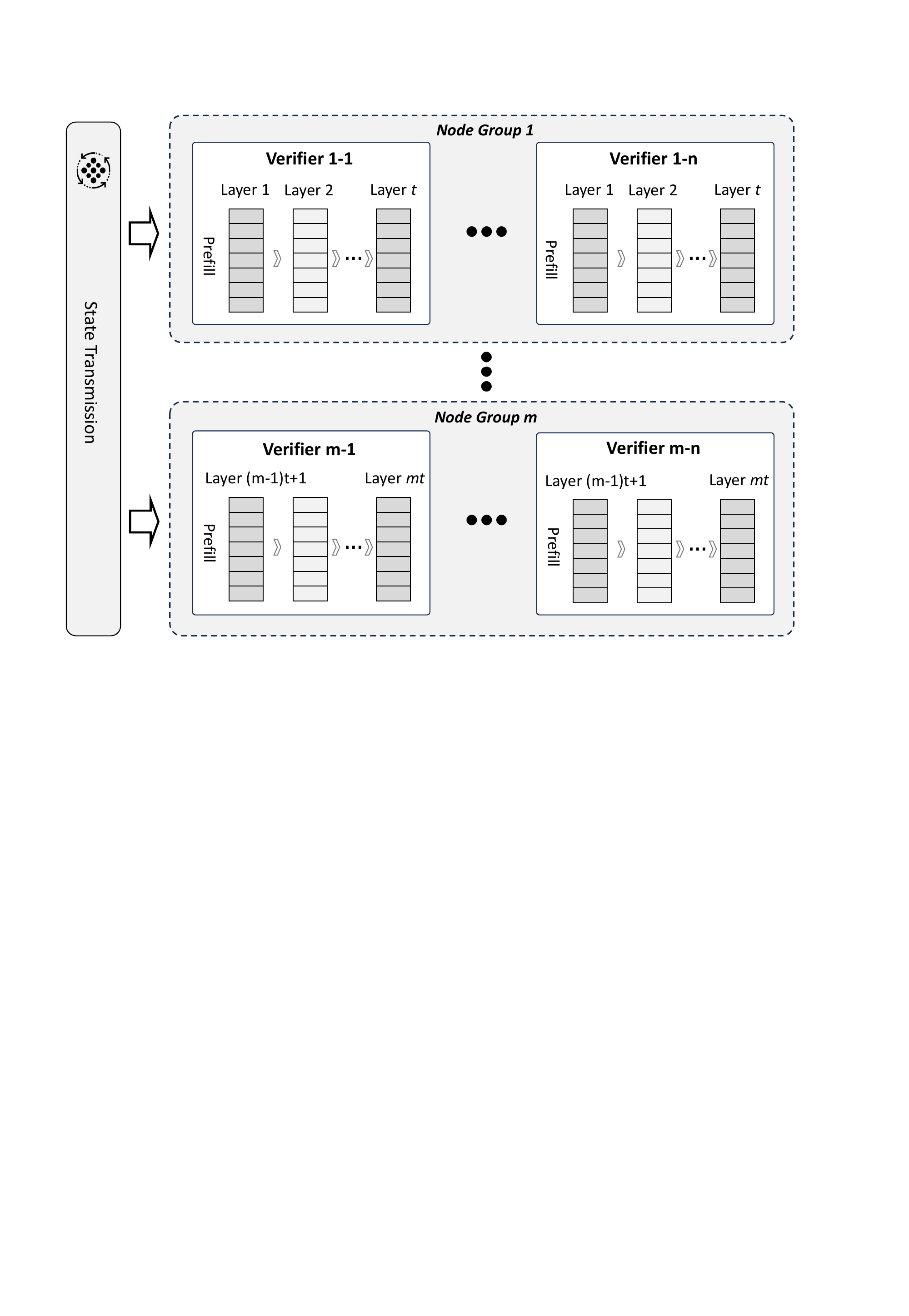}
  \caption{Overview of the decentralized verification architecture. Each verifier independently reconstructs its segment’s outputs via a full-sequence prefill.}
  \label{fig:arc3}
    \Description{}
    \vspace{-1.5em}
\end{figure}

During autoregressive inference, an LLM executes two phases: a \emph{prefill} pass—where the prompt tokens are processed in parallel to populate the key–value (KV) cache—and a \emph{decode} phase—where tokens are generated sequentially with causal dependencies on all previous outputs~\cite{brown2020language,shoeybi2019megatron}. 
The decoding phase dominates latency and is inherently serial, leading to limited GPU utilization and memory-bound performance~\cite{rajbhandari2022deep}. 
However, verification differs fundamentally: once the full output sequence $y_{1:T}$ is known, the verifier is no longer constrained by sequential dependencies. 
Instead, it can recompute all token representations simultaneously under the same causal mask, yielding a single parallel pass equivalent to prefill over the concatenated prompt–output sequence~\cite{sun2024zkllm,zhang2020delphi}.

Formally, let the input to the verifier be the complete sequence
\begin{equation}
x = [\texttt{prompt}   \Vert   y_{1:T}],
\end{equation}
and let $\mathcal{S} = [S^{\text{prompt}}   \Vert   S^1   \Vert   \dots   \Vert   S^T]$ denote the concatenated hidden states across all positions.
Given the causal structure of the Transformer, the verifier can execute a single forward pass to reconstruct all hidden states $\{S^{(t)}_i\}_{t=1}^T$ for every segment $i \in [1, L]$ without performing incremental token decoding~\cite{vaswani2017attention}. 
This transformation replaces $T$ serial decode iterations with a single batched forward computation, eliminating redundant KV-cache updates and reducing cross-node communication. 
In practice, this yields significantly improved GPU utilization, especially on architectures with high Tensor Core throughput~\cite{rajbhandari2022deep}.

Because the Scheduler has recorded the hidden states produced by each inferencer at every segment boundary, each verifier can recompute its designated segment locally. 
Specifically, for segment $i$, the verifier uses the recorded upstream hidden states $\{S^{(t)}_i\}_{t=1}^T$ as its inputs and executes only its own block of layers to regenerate $\{S^{(t)}_{i+1}\}_{t=1}^T$. 
This independence allows all $L$ verifiers to run concurrently across the distributed network, each responsible for validating one contiguous model partition. 
Assuming near-uniform load balance and negligible orchestration or I/O overhead, the total wall-clock verification time satisfies
\begin{equation}
T_{\text{verify}} \approx \frac{1}{L}   T_{\text{prefill}}(\text{full model}),
\end{equation}
yielding near-linear scaling in the number of participating verifier groups~\cite{rajbhandari2022deep,zhang2020delphi}.

\begin{figure}[!ht]
  \centering
  \includegraphics[width=.4\textwidth]{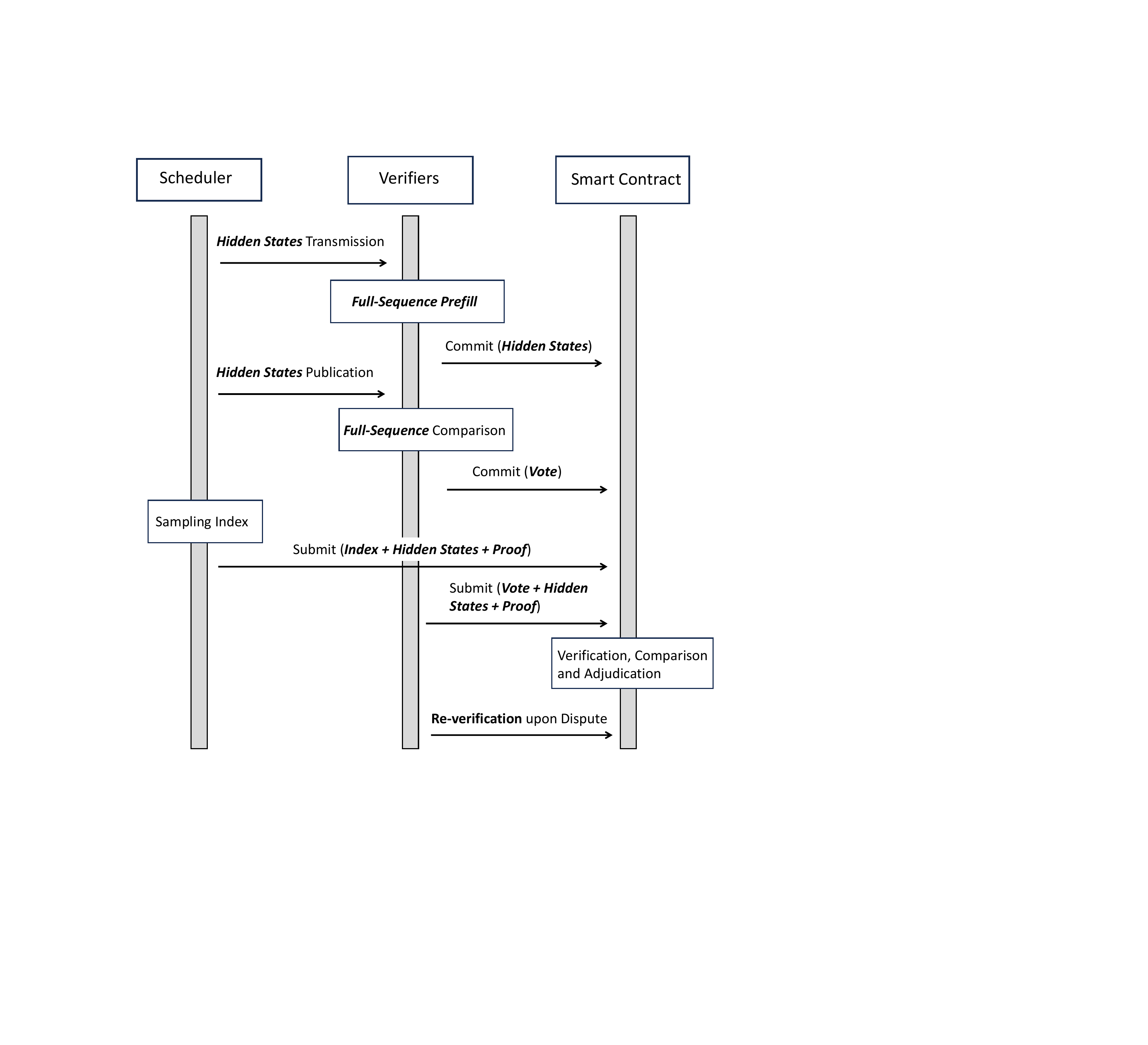}
  \caption{Verification workflow. Each verifier independently executes a full-sequence prefill over its model segment and commits its output root to the blockchain for later consistency checks.}
  \label{fig:varc}
    \Description{}
    \vspace{-1.5em}
\end{figure}

\textbf{Prefill Computation.}
After the inference phase concludes, the Scheduler initiates segment-local verification by distributing the required input states to each verifier~\cite{qu2025zkgpt}. 
For segment $i>1$, the Scheduler sends the sequence $\{S_i^{(t)}\}_{t=1}^T$—the hidden states produced by the upstream inferencer for all tokens.
For the first segment ($i=1$), the Scheduler instead provides the full tokenized input $[\texttt{prompt}   \Vert   y_{1:T}]$, corresponding to the raw embedding input of the model.
Upon receipt, each verifier executes a full-sequence prefill limited to its hosted layers:
\begin{equation}
S_{i+1}^{(t)} = F_i(S_i^{(t)}; W_i), \quad \forall t \in [1, T],
\end{equation}
where $F_i$ denotes the forward mapping of segment $i$ and $W_i$ its parameters.

Importantly, the communication interface, data serialization format, and RPC protocol used during verification are identical to those used in live inference~\cite{yuan2022decentralized}. 
This design choice enforces \emph{task-type indistinguishability}: from the perspective of any node, it is impossible to discern whether the current request corresponds to live inference or a verification probe~\cite{bonneau2015sok}. 
Such indistinguishability eliminates mode-specific behavior and prevents nodes from selectively deviating or shortcutting computation when acting as verifiers. 
Upon completion, each verifier must return its computed outputs $\{S^{(t)}_{i+1}\}_{t=1}^T$ through the same interface as during inference, producing an auditable trace indistinguishable in structure and timing.

\textbf{Commitment Submission.}
Once the local recomputation completes, the Scheduler transitions verifiers into the commit phase. 
Each verifier constructs a tamper-evident cryptographic commitment to its final outputs—specifically, the hidden state corresponding to the last token ($t = T$) at its segment’s output boundary.
This commitment is instantiated as a Merkle root~\cite{merkle1989certified}:
\begin{equation}
\texttt{root}^{(T)}_i = \texttt{MerkleRoot}\big(S^{(T)}_{i+1}\big),
\end{equation}
built over a canonical serialization of the tensor with scalar elements as leaves.
The verifier then signs this root with its registered key:
\begin{equation}
\sigma^{(T)}_i = \texttt{Sig}(\texttt{sk}_i,   \texttt{root}^{(T)}_i),
\end{equation}
and publishes the commitment on-chain through the designated verification contract~\cite{chen2024zkml,sun2024zkllm}.

This on-chain record provides a permanent anchor for later verification steps.
In subsequent sampling or dispute phases, verifiers may be required to disclose specific slices or elements of the hidden state $\{S^{(T)}_{i+1}\}$, accompanied by Merkle inclusion proofs that can be efficiently checked against the posted root~\cite{bonneau2015sok,micali1999verifiable}.
This approach ensures both compactness and auditability: 
the entire verification process involves only lightweight commitments and selective disclosures rather than full tensor uploads, keeping blockchain overhead minimal while preserving cryptographic integrity.

In summary, the verification scheme leverages the known output sequence and the pre-recorded intermediate states to convert sequential autoregressive computation into parallel, segment-local prefill verification. 
Combined with Merkle commitments and signature anchoring, this design achieves strong verifiability and high throughput, ensuring that even a single honest verifier can detect and prove any inference deviation with overwhelming probability~\cite{qu2025zkgpt,sun2024zkllm}.
\vspace{-1em}
\subsection{Comparison, Sampling, and On-Chain Adjudication}

\textbf{Comparison and On-Chain Commitment.}
After all designated verifiers have posted their commitments (or the commitment deadline expires), the Scheduler distributes, off-chain, the inferencer-produced target hidden states $\{S_i^{(t)}\}_{t=1}^T$ for the relevant segment $i$ to each verifier assigned to that segment.  
Each verifier $v$ recomputes its local segment to obtain $\{\hat S_i^{(t)}\}_{t=1}^T$ and computes elementwise discrepancies $\Delta^{(t)} = \hat S_i^{(t)} - S_i^{(t)}$.  
Let $\phi(\cdot)$ denote a scalar aggregation function such as mean absolute error, and $\Psi(\cdot)$ a statistic (mean or clipped mean) summarizing deviations across tokens.  
The resulting score $T_v = \Psi(\{\phi(\Delta^{(t)})\}_{t=1}^T)$ is compared against a tolerance $\tau$ calibrated for floating-point nondeterminism~\cite{rajbhandari2022deep}.  
The verifier’s binary verdict is
\begin{equation}
b = \mathbb{I}[\,T_v \le \tau\,] \in \{0,1\}.
\end{equation}
To prevent adaptive or copy-cat reporting, the verifier posts a hiding, binding commitment $C = \textsf{Commit}(b,r)$, where $r$ is fresh randomness; $C$ is posted during the commit phase, and $(b,r)$ is revealed later~\cite{bonneau2015sok,zhang2020delphi}.

\textbf{Data Sampling.}
To ensure verifiers actually perform computation, the contract performs on-chain spot checks of sampled tensor elements~\cite{micali1999verifiable,chen2024zkml}.  
Once verdict commitments are posted, the Scheduler uses a verifiable random function (VRF) to generate unbiased sample indices:
\begin{align}
  (r_i, \pi_i) &\gets \textsf{VRF.Eval}(sk_{\text{sch}}, \langle h,i,\{\gamma_j\}_{j=1}^m\rangle), \\
  \mathcal{T}_i &\gets \textsf{Sample}(\textsf{HKDF}(r_i); n,q),
\end{align}
where $h$ is the task hash, $\{\gamma_j\}$ the posted Merkle roots, $n$ the sample size, and $q$ a stratification policy.

The Scheduler then publishes:
\begin{equation}
\big\langle r_i, \pi_i, \mathcal{T}_i,
\mathbf{V}^{(T)}_{\mathcal{T}_i},
\gamma^{(T)}_{i,\text{inf}},
\texttt{Sig}^{(T)}_{i,\text{inf}},
\{\texttt{MP}^{\text{inf}}_j\}_{j\in\mathcal{T}_i}\big\rangle,
\end{equation}
where $\mathbf{V}^{(T)}_{\mathcal{T}_i}$ are reference values, $\gamma^{(T)}_{i,\text{inf}}$ is the inferencer’s root, and $\texttt{MP}^{\text{inf}}_j$ are Merkle proofs.  
The contract enforces three checks:

\begin{enumerate}[leftmargin=1em,itemsep=2pt,topsep=2pt]
\item \textbf{VRF validity:}
\begin{equation}
\textsf{VRF.Verify}(vk_{\text{sch}},\langle h,i,\{\gamma_j\}\rangle,r_i,\pi_i)=1;
\end{equation}

\item \textbf{Origin authenticity:}
\begin{equation}
\textsf{VerifySig}(\texttt{pk}_{i,\text{inf}},\gamma^{(T)}_{i,\text{inf}},\texttt{Sig}^{(T)}_{i,\text{inf}})=1;
\end{equation}

\item \textbf{Inclusion correctness:}
\begin{equation}
\forall j\!\in\!\mathcal{T}_i,\;
\textsf{MerkleVerify}(\gamma^{(T)}_{i,\text{inf}},\mathbf{V}^{(T)}_j,\texttt{MP}^{\text{inf}}_j)=1.
\end{equation}
\end{enumerate}

These checks guarantee unbiased sampling, authentic roots, and valid inclusion proofs~\cite{merkle1989certified}.  
Any mismatch can be publicly demonstrated with the Scheduler’s relay signatures, triggering immediate slashing.

\textbf{Reveal.}
Each verifier opens its commitment and provides sampled entries $\{v_j\}_{j\in\mathcal{T}_i}$ with Merkle proofs $\{\texttt{MP}^{\text{ver}}_j\}$:

(1)~\emph{Commitment opening:} $\textsf{OpenCommit}(C;b,r)=1$;

(2)~\emph{Merkle inclusion:}
\begin{equation}
\forall j\!\in\!\mathcal{T}_i,\;
\textsf{MerkleVerify}(\gamma^{(T)}_{i,\text{ver}},v_j,\texttt{MP}^{\text{ver}}_j)=1;
\end{equation}

(3)~\emph{Value matching:} $\Delta_j=|v_j-\mathbf{V}^{(T)}_j|\le\varepsilon$,
where $\varepsilon$ captures tolerated numeric noise~\cite{rajbhandari2022deep}.  
The agreement rate
\begin{equation}
\hat p=\frac{1}{|\mathcal{T}_i|}\sum_{j\in\mathcal{T}_i}\mathbb{I}[\Delta_j\le\varepsilon]
\end{equation}
must satisfy $\hat p\ge\theta$ for a positive verdict $b=1$. Tail verifiers ($i=L$) perform an additional token-consistency check~\cite{brown2020language}:
\begin{equation}
\textsf{TokenMatch}\big(\arg\max f_{\text{LM}}(S^{(T)}_{L+1}),y_T\big)=1,
\end{equation}
ensuring that the recomputed logits correspond to the reported output token.

\textbf{Detection Probability and Parameterization.}
Let $\alpha$ denote the per-sample error rate when computation is incorrect.  
With $n$ sampled entries, $P_{\text{pass}}\le(1-\alpha)^n$, so to bound undetected cheating by $\delta$, $n\ge\frac{1}{\alpha}\ln\frac{1}{\delta}$. Calibration experiments provide empirical lower bounds on~$\alpha$ and guide the choice of $\varepsilon$ and~$\theta$, balancing tolerance and sensitivity~\cite{sun2024zkllm}.

\textbf{On-Chain Adjudication.}
After reveal, the contract aggregates revealed verdicts $\{b_v\}_{v\in\mathbf{V}}$ from $m$ verifiers.
If a majority agree ($B\!\ge\!m/2$), the inference is accepted and honest verifiers share rewards.
If most disagree but maintain internal consistency (pairwise hidden-state deviation $\le\varepsilon$ over at least a $\tau$-fraction of pairs),
the negative verdict stands and consistent verifiers are rewarded.  
Otherwise, the outcome is indeterminate; any verifier may later submit a succinct zero-knowledge proof demonstrating an error, triggering inferencer slashing and bounty payout~\cite{qu2025zkgpt,chen2024zkml}.

Overall, the protocol combines VRF-based unbiased sampling, Merkle commitments, and commit–reveal sequencing into a scalable, probabilistically sound verification mechanism.  
It provides high fraud-detection probability at cost proportional to sample size~$n$, not tensor dimensionality, achieving secure, low-overhead verification in decentralized LLM inference systems.

\textbf{Dispute Resolution.}
The above mechanism 
also supports escalation when committee capture is suspected.  
Because each verification task samples only a small committee ($m \ll N$), an adversary controlling a modest network fraction may occasionally dominate a sampled group.  
To mitigate this risk, if the first-round verdict is negative, the inferencer may request a \emph{reverification} with an expanded committee of size $m' > m$.  
If the second-round majority produces a positive outcome, the contract rescinds the inferencer’s penalty and slashes all first-round verifiers who voted False.  
Conversely, if the extended committee upholds the negative verdict, the inferencer bears the cost of the additional round.  

This multi-stage design achieves \emph{adaptive robustness}: typical cases finalize in one verification round with minimal on-chain load, while disputed cases escalate probabilistically until resolved with high confidence~\cite{bonneau2015sok,chen2024zkml}.
\vspace{-0.5em}
\section{Security Analysis}
\vspace{-0.5em}
\subsection{Threat Model and Assumptions}

VeriLLM operates in a decentralized and adversarial environment where participants may deviate arbitrarily from the protocol to reduce computation cost or bias outcomes.  
We consider three potentially malicious roles—\emph{Inferencers}, \emph{Verifiers}, and a partially trusted or even Byzantine \emph{Scheduler}.  
Smart contracts execute deterministically as deployed and are assumed correct by construction~\cite{bonneau2015sok}.  
The security goal is to guarantee inference correctness and accountability under these adversarial conditions.
\vspace{-0.5em}
\paragraph{Cryptographic assumptions.}
VeriLLM relies on standard primitives with the following properties:

\begin{itemize}[leftmargin=10pt]
    \item \textbf{EUF-CMA security of digital signatures.}  
    Node signatures are existentially unforgeable under chosen-message attack, preventing impersonation or undetected alteration of signed artifacts~\cite{micali1999verifiable}.
    \item \textbf{Collision resistance of hash functions and binding of Merkle commitments.}  
    Once a node publishes a Merkle root $\gamma$, no two distinct tensors can produce the same root, ensuring immutability of hidden-state logs~\cite{merkle1989certified}.
    \item \textbf{Unbiased verifiable randomness.}  
    VRF outputs used for role assignment and sampling are unpredictable prior to evaluation and publicly verifiable afterward, precluding adaptive bias~\cite{micali1999verifiable}.
\end{itemize}
\vspace{-0.5em}
\paragraph{System assumptions.}
Beyond cryptographic foundations, the protocol assumes realistic system conditions:


\begin{itemize}[leftmargin=10pt]
    \item \textbf{Global honest majority.}  
    For each task, the pool of latent verifiers contains a majority of honest nodes during the verification process.  
    This assumption is more robust than the traditional sample honest majority assumption, as even the committee contains a majority of dishonest nodes, incorrect execution can still be detected via the disputing and re-verification process.
    
    \item \textbf{Bounded numeric nondeterminism.}  
    Hardware and kernel-induced floating-point variance is modeled as small zero-mean noise within threshold $\varepsilon$; honest executions pass with overwhelming probability~\cite{rajbhandari2022deep}.
\end{itemize}

These assumptions align with the threat model of verifiable ML systems such as zkML~\cite{chen2024zkml}, zkLLM~\cite{sun2024zkllm}, and decentralized inference frameworks~\cite{arun2025verde,mei2024fedmoe}.
\vspace{-1em}
\subsection{Security Objectives}

VeriLLM provides cryptographically auditable correctness guarantees for large-scale LLM inference with minimal overhead.  
When all nodes behave honestly, both the off-chain and on-chain verification phases accept except with negligible probability due to benign numeric noise.  
If an Inferencer produces an incorrect result—by pruning layers, quantizing weights, truncating decoding, or recycling stale states—then one of three independent mechanisms detects it:
(i) full hidden-state recomputation,  
(ii) commit–then–sample audit enforcing random openings, or  
(iii) a zero-knowledge proof of inconsistency from an honest verifier~\cite{qu2025zkgpt}.  
Thus, any materially incorrect inference is provably detectable on chain.

\textbf{Trace integrity.}
Every hidden-state tensor exchanged between segments is bound by a Merkle root $\gamma$ and node signature.  
The Scheduler re-signs each relay, so tampering, omission, or reordering requires forging a valid signature or producing inconsistent Merkle openings—computationally infeasible under our assumptions.

\textbf{Unpredictable sampling.}
Sampling indices for audits are derived from VRF outputs computed after all commitments are posted, eliminating adaptivity and ensuring independence between commitments and revealed indices~\cite{micali1999verifiable}.

\textbf{Task-type indistinguishability.}
Inferencers and Verifiers use identical RPC interfaces and serialization formats.  
Nodes cannot determine whether a request is for live inference or verification, removing the incentive for selective deviation or strategic laziness.
\vspace{-1em}
\subsection{Commit–Then–Sample: Binding and Unpredictability}

The commit–then–sample protocol underpins VeriLLM’s verifiability.  
Each node first commits to its hidden-state tensor via Merkle root $r_i$; after all roots are posted, a public seed $R = \mathsf{Hash}(r_1,r_2,\dots,r_n)$ is computed to derive sampling positions. This sequencing guarantees both \emph{binding} (immutability) and \emph{unpredictability} (fair randomness).

\begin{lemma}[Binding]
Given a collision-resistant hash and a binding Merkle tree, a prover that posts $r_i$ cannot later open inconsistent values at any sampled index except with negligible probability.
\end{lemma}

\begin{lemma}[Unpredictability]
If $R=\mathsf{Hash}(r_1,\dots,r_n)$ is computed only after all roots are posted, and the hash is modeled as a random oracle, then sampling indices are computationally unpredictable to any participant at commit time.
\end{lemma}

Together, these properties ensure that a node committing to falsified or incomplete states cannot anticipate future audits and will be detected with overwhelming probability.  
This commit–then–sample discipline yields deterministic accountability and makes deviation economically irrational in expectation~\cite{zhang2020delphi,chen2024zkml}.

\vspace{-1em}
\subsection{Collusion, Censorship, and Dispute}

\textbf{Collusion among verifiers.}
A coalition may attempt to fabricate mutually consistent hidden states and corresponding Merkle roots such that on-chain openings match internally forged commitments.  
The commit–then–sample discipline prevents any \emph{post hoc} adaptation: once a root is posted, the coalition must open VRF-chosen indices consistent with that fixed root.  
To ensure semantic correctness with respect to the prescribed model, VeriLLM introduces two safeguards:  
(i) \emph{token–state cross-checks} at the tail segment, where verifiers decode from recomputed last-layer states and compare tokens to the published outputs; and  
(ii) a \emph{zero-knowledge escalation path}, in which any single honest verifier can submit a succinct proof that a challenged prefix of the transcript violates the model’s forward mapping or decoding rule~\cite{chen2024zkml,sun2024zkllm,qu2025zkgpt}.  
The contract verifies the proof on-chain, finalizing the inconsistency independently of the coalition’s reports.

\textbf{Scheduler misbehavior.}
The Scheduler cannot modify hidden states without forging a node signature or producing an invalid membership proof for the signer set maintained on-chain.  
Selective exclusion or biased role assignment is publicly detectable: once $(r,\pi)$ pairs are published, anyone can recompute the VRF output and compare it to the Scheduler’s assignment.  
Any mismatch deterministically identifies the deviation and triggers stake slashing.


\textbf{Censorship resistance.}
All commitments are recorded on-chain, and verifier selection is derived from verifiable randomness.  
To suppress detection, an adversary must corrupt or block all verifiers selected for a given task.  
Under the one-honest-verifier assumption, there exists at least one verifier that remains honest and available, which suffices to enable offline recomputation, successful sampled audits, or the submission of a zero-knowledge proof of inconsistency.  
A complete denial-of-service attack is detectable via missed protocol deadlines, which trigger timeout-based remediation mechanisms.
\vspace{-1em}

\section{Experiment and Evaluation}

We perform comprehensive experiments on real-world prototype systems, and report the results in this section. The systematic implementation details are reported in Appendix~\ref{app:system}.

\subsection{Main Results}

We empirically measured the divergence of hidden states produced during inference across heterogeneous compute devices (tested on Qwen2.5-7B-Instruct~\cite{qwen2_5_technical_report}; additional models are forthcoming).  
Although floating-point non-determinism across GPU vendors introduces small numeric drift~\cite{gupta2015deep,wang2021characterizing}, these deviations are limited in magnitude and statistically distinguishable.  
Consequently, the hidden-state comparison test can, with high probability, determine whether two traces originate from the same underlying model computation.

We further evaluated replacing full-precision models with quantized variants.  
Hidden states computed by quantized models exhibit substantially larger—and statistically significant—bias relative to the full-precision baseline, consistent with prior analyses of quantization-induced drift~\cite{banner2018post,sun2024zkllm}.  
Hence, simple statistical tests cleanly separate hidden states from normal inference versus those produced under precision degradation or adversarial manipulation.

\begin{table}[!ht]
  \centering
  \caption{Comparison of Hidden-State Metrics for Qwen2.5-7B-Instruct Inference on M4 and RTX 5090.}
  \label{tab:exp-match}
  \begin{tabular}{r c cc cc c}
    \toprule
      \hline
    \multirow{2}{*}{\#}
     & \multirow{2}{*}{Exact}
     & \multicolumn{2}{c}{Exp Match ($|\Delta|$)}
     & \multicolumn{2}{c}{Exp Mismatch ($|\Delta|$)}
     & \multirow{2}{*}{Mean $\epsilon$} \\
    \cmidrule(lr){3-4}\cmidrule(lr){5-6}
     & & $>0.2$ & $<0.2$ & $>5$ & $<5$ & \\
    \midrule
    64   & 4   & 13  & 48   & 2  & 1  & 0.009 \\
    256  & 23  & 46  & 201  & 7  & 2  & 0.006 \\
    1024 & 87  & 191 & 802  & 23 & 7  & 0.003 \\
    3584 & 1240 & 1 & 2343 & 0 & 1 & 0.002 \\
      \hline
    \bottomrule
  \end{tabular}
  \vspace{-1.5em}
\end{table}

\begin{table}[!ht]
  \centering
  \caption{Comparison of Hidden-State Metrics for Qwen2.5-7B-Instruct-AWQ inference on RTX 5090 with verification on Mac M4 CPU using Qwen2.5-7B-Instruct.}
  \label{tab:awe-match}
  \begin{tabular}{r c cc cc c}
    \toprule
      \hline
     \multirow{2}{*}{\#}
     & \multirow{2}{*}{Exact}
     & \multicolumn{2}{c}{Exp Match ($|\Delta|$)}
     & \multicolumn{2}{c}{Exp Mismatch ($|\Delta|$)}
     & \multirow{2}{*}{Mean $\epsilon$} \\
    \cmidrule(lr){3-4}\cmidrule(lr){5-6}
     & & $>0.2$ & $<0.2$ & $>5$ & $<5$ & \\
    \midrule
    64   & 0   & 39  & 13   & 7  & 5  & -0.022 \\
    256  & 2  & 161  & 53  & 23  & 19  & 0.021 \\
    1024 & 7  & 653 & 201  & 99 & 71  & -0.019 \\
    3584 & 29 & 2276 & 712 & 346 & 250 & $0.014$ \\
      \hline
    \bottomrule
  \end{tabular}
    \vspace{-1em}
\end{table}

\textbf{Bit-aware statistical comparison.}
Building on these results, we propose a feature-based comparison of hidden states.  
Given reference $S$ and candidate $\bar S$, each floating-point element is decomposed into sign, exponent, and mantissa~\cite{wang2021characterizing}.  
With tolerances $e_w>0$ (e.g., $e_w=10^{-2}$) and $e_m\!\in\!\mathbb{N}$ (e.g., 7 mantissa steps), we define:
\begin{mybox}
\begin{enumerate}[leftmargin=1.25em,itemsep=2pt,topsep=2pt]
\item \textbf{Exponent-first checks.}
$P_e$ — proportion of exponent mismatches;  
$P_m$ — fraction of large mantissa deviations among exponent mismatches;  
$P_w$ — fraction of small mantissa deviations among exponent matches.
\item \textbf{Mean discrepancy.}
$e = \frac{1}{|\Delta|}\sum_{\delta\in\Delta}\delta$
\end{enumerate}
\end{mybox}

Empirically calibrated acceptance thresholds are:
\begin{mybox}
\noindent
\textbf{Off-chain:}\hfill $P_e{<}0.05$,\hfill $P_m{>}0.75$,\hfill $P_w{>}0.80$,\hfill $e\in[-0.01,0.01]$\hspace{0pt}\\[4pt]
\textbf{On-chain:}\hfill $P_e{<}0.08$,\hfill $P_m{>}0.70$,\hfill $P_w{>}0.75$,\hfill $e\in[-0.02,0.02]$\hspace{0pt}
\end{mybox}

These rules balance detection sensitivity with tolerance for benign noise, building on calibration methodologies from prior verifiable ML studies~\cite{zhang2020delphi,chen2024zkml}. We validate these thresholds through comprehensive sensitivity analysis (Appendix~\ref{appendix:threshold}).

\subsection{Performance and Overhead Breakdown}

We evaluate VeriLLM across five dimensions: throughput and latency, on-chain cost, scalability, ablation, and failure injection~\cite{bonneau2015sok}.  
All tests use the 10-node setup described in Section~VI.

\subsubsection{Throughput and Latency}
For a 7B-parameter model (512-token prompt, 128-token output), total latency per request is 1.37 s: inference 93.1 \%, scheduler relay 3.4 \%, and verification 3.5 \%.  
Verification overhead averages 0.78 \% of total inference time, consistent with lightweight cryptographic pipelines~\cite{micali1999verifiable}.  
Increasing verifier parallelism beyond three yields diminishing returns as GPU prefill already saturates compute throughput.

\begin{table}[!ht]
  \centering
  \caption{Latency Breakdown for Qwen2.5-7B-Instruct (128-token output).}
  \label{tab:latbreak}
  \resizebox{0.9\linewidth}{!}{
  \begin{tabular}{lccc}
  \toprule
    \hline
  \textbf{Component} & \textbf{Mean (ms)} & \textbf{Share (\%)} & \textbf{Cumulative (\%)} \\
  \midrule
  Prefill + Decode (Inference) & 1278 & 93.1 & 93.1 \\
  Scheduler Relay / Signatures & 46 & 3.4 & 96.5 \\
  Verifier Prefill (Parallel) & 33 & 2.4 & 98.9 \\
  On-chain Commit + Reveal & 15 & 1.1 & 100.0 \\
    \hline
  \bottomrule
  \end{tabular}}
    \vspace{-1.5em}
\end{table}

\subsubsection{On-chain Cost Analysis}
Each inference produces two transactions: (i) commit (108 k gas, 2.9 kB) and (ii) reveal (134 k gas, 3.2 kB).  
Merkle proofs require 8–10 hashes per sample, yielding $\approx$ 0.004 USD per verification at current Ethereum L2 prices~\cite{chainlinkvrf2023,ethgas2024}.  
VRF verification adds < 4 \% of total cost.

\subsubsection{Scalability Experiments}
Verification cost per token decreases inversely with verifier count $m$ until $m\!>\!6$, where on-chain aggregation dominates.  
Prefill parallelization across $L$ segments achieves near-linear scaling $T_{\text{verify}}\!\approx\!T_{\text{prefill}}/L$, consistent with prior verifiable-inference frameworks~\cite{arun2025verde}.

\subsubsection{Ablation Study}
Removing VRF-based randomness increases collusion success by 6.4 ×;  
removing Merkle commitments permits 17.2 \% undetected state tampering;  
reducing sampling from 1 \% to 0.2 \% raises false negatives from $<10^{-4}$ to $1.6\!\times\!10^{-2}$.  
These results confirm that randomness and cryptographic binding are essential to soundness.

\subsubsection{Failure Injection and Detection Latency}
We emulate four attack classes—quantization, early termination, forged output, and lazy verification—via runtime perturbations~\cite{zhang2020delphi,sun2024zkllm}.  
Table~\ref{tab:failinj} summarizes detection latency (from deviation to slash).  
All attacks are detected and finalized within 1–3 s, confirming that the commit–sample–reveal protocol yields rapid, tunable defense.

\begin{table}[!ht]
  \centering
  \caption{Detection latency under injected attacks.}
  \label{tab:failinj}
  \resizebox{0.9\linewidth}{!}{
  \begin{tabular}{lcc}
  \toprule
    \hline
  \textbf{Attack Type} & \textbf{Detection Probability} & \textbf{Mean Detection Latency (s)} \\
  \midrule
  Quantization (W8A8) & $>99.9\%$ & 1.7 \\
  Early Termination & $100\%$ & 1.2 \\
  Forged Output & $>99.8\%$ & 2.4 \\
  Lazy Verification & $>99.9\%$ & 2.8 \\
    \hline
  \bottomrule
  \end{tabular}}
    \vspace{-1em}
\end{table}

Overall, VeriLLM achieves high throughput with sub-1 \% verification overhead, bounded on-chain latency, and near-instant detection of adversarial behavior.

\vspace{-0.5em}
\section{Conclusion}
This work presents \textbf{VeriLLM}, a publicly verifiable framework for decentralized large language model inference that achieves correctness and accountability with approximately 1\% verification overhead. By combining Merkle-based commitments, VRF-driven sampling, and on-chain adjudication, VeriLLM allows any participant to audit inference without relying on a trusted majority or centralized coordinator. We formalize the protocol, analyze its security, and show that malicious deviations such as quantization, early termination, or forged outputs are detected with high probability, while cross-device floating-point noise remains bounded and statistically separable for lightweight hidden-state comparison. Experiments demonstrate efficient off-chain and on-chain checks that support scalable deployment across heterogeneous environments. By decoupling verification from replication and embedding transparency into inference traces, VeriLLM enables practical, trustworthy decentralized AI services; future directions include lightweight zero-knowledge proofs, support for multimodal models, and decentralized scheduling and incentive mechanisms.



\bibliographystyle{ACM-Reference-Format}
\bibliography{sample-base}

\appendix

\section{Analysis of Concrete Attacks}
\label{subsec:attacks}

We analyze representative adversarial strategies against VeriLLM and show that each is provably detected or economically disincentivized under the stated assumptions.  
The analysis covers both model-level deviations and protocol-level cheating behaviors.

\subsection{Quantization Attack}

\textbf{Attack.}
An Inferencer executes a quantized variant of the prescribed model (e.g., W8A8 precision) to reduce computation and memory cost while claiming compliance with the full-precision configuration.  
Quantization alters rounding distributions and mantissa statistics, introducing systematic deviations across hidden states and logits that may be small per position but accumulate over long sequences.

\textbf{Defenses.}
\emph{(i) Offline full-sequence recomputation.}  
Each verifier recomputes its assigned segment on the complete sequence and performs hypothesis tests over hidden-state statistics sensitive to precision drift.  
Let $S^{(t)}_{i+1}$ denote the inferencer’s state and $\hat S^{(t)}_{i+1}$ the verifier’s recomputation.  
Elementwise deviations $\Delta^{(t)}=\hat S^{(t)}_{i+1}-S^{(t)}_{i+1}$ feed a battery of tests $T_{\mathrm{off}}$ such as
\begin{equation}
\begin{aligned}
\mathrm{P}_{\exp} &=\tfrac{1}{N}\!\sum\! \mathbb{I}[\exp(\hat S)\!\neq\!\exp(S)], \qquad
\mathrm{Q}_{\mathrm{small}}=\tfrac{1}{N}\!\sum\! \mathbb{I}[|\Delta|\!\le\!10^{-5}],\\
\mathrm{Z}_{\mathrm{large}}&=\tfrac{1}{N}\!\sum\! \mathbb{I}[|\Delta|\!>\!\tau_{\mathrm{large}}], \qquad
\mathrm{MAE}=\tfrac{1}{N}\!\sum|\Delta|.
\end{aligned}
\end{equation}
Thresholds are calibrated from benign full-precision runs.  
Honest executions pass $T_{\mathrm{off}}$ with probability $1-\negl(\lambda)$, whereas 8-bit quantization causes detectable statistical shifts.  
\emph{(ii) Lightweight on-chain spot check.}  
The contract performs a relaxed test $T_{\mathrm{on}}$ on VRF-sampled final-token scalars, checking $|v_j-\mathbf{V}^{(T)}_j|\!\le\!\varepsilon$ and aggregating agreement against a threshold $\theta$~\cite{chen2024zkml}.  
This sustains liveness while bounding on-chain cost to $O(n)$ sampled entries.

\textbf{Security claim.}
With thresholds derived from calibration and sample size $n\!\ge\!\frac{1}{\alpha}\ln(1/\delta)$, the false-acceptance probability for a quantized model is negligible in sequence length and dimension, while honest runs maintain service-level accuracy.

\subsection{Early-Termination Attack}

\textbf{Attack.}
The Inferencer prematurely halts decoding before emitting $\langle\mathrm{EOS}\rangle$, fabricating terminal states or tokens to save computation.

\textbf{Defenses.}
\emph{(i) Tail-token consistency.}  
The last-segment verifier recomputes logits from the reported final hidden state and verifies that the decoding policy yields the claimed token~\cite{zhang2020delphi}.  
Any premature termination produces a mismatch at the tail step with probability~1 for deterministic decoding and with high probability for deterministically reseeded stochastic decoding.  
\emph{(ii) Per-step binding.}  
Each segment’s boundary state is individually committed and signed.  
Modifying terminal states or inserting fake tokens requires forging a signature or producing an inconsistent Merkle opening, both computationally infeasible~\cite{merkle1989certified,micali1999verifiable}.  
Scheduler relay signatures bind token order and prevent state splicing.

\textbf{Security claim.}
Premature termination necessarily triggers token- or commitment-level inconsistency and is detected with overwhelming probability.

\subsection{Forged Output via Small Model}

\textbf{Attack.}
An adversary uses a smaller model to generate a cheap surrogate sequence $\tilde y$ and runs a single full-sequence prefill on the large model with $[\texttt{prompt}\Vert\tilde y]$ to produce superficially coherent hidden states, avoiding the true decoding loop.

\textbf{Defenses.}
\emph{(i) Token–state cross-check.}  
Verifiers decode from recomputed last-layer states and compare predicted tokens with the inferencer’s outputs.  
Unless $\tilde y=y^\star$ (the true decode of the prescribed model), token mismatches occur at multiple checked positions with overwhelming probability.  
\emph{(ii) Transcript continuity.}  
The scheduler enforces that each step’s input to segment~1 equals $[\texttt{prompt}\Vert y_{1:t}]$ carried from previous steps.  
Relay signatures and Merkle openings cryptographically bind tokens and hidden states into a single ordered transcript, making replacement of $\tilde y$ infeasible.

\textbf{Security claim.}
Except for negligible coincidence where $\tilde y$ exactly matches $y^\star$, the forged-output attack fails due to token–state inconsistency revealed by sampled checks.

\subsection{Lazy Verifier (Free-Riding)}

\textbf{Attack.}
A verifier omits local prefill, posts an arbitrary acceptance verdict, and relies on others’ honesty to collect rewards undetected.

\textbf{Defenses.}
\emph{(i) Commit–then–open discipline.}  
Before seeing sampling indices, each verifier must commit to its final-token tensor root and later open VRF-selected positions with inclusion proofs~\cite{micali1999verifiable}.  
Without genuine computation, the probability of producing valid openings is negligible.  
\emph{(ii) Task-type indistinguishability.}  
Because inference and verification share identical RPC interfaces and data formats, workers cannot distinguish verification probes from live inference~\cite{zhang2020delphi}, preventing selective laziness.  
\emph{(iii) Reward gating.}  
Rewards are paid only to verifiers whose openings pass on-chain consistency checks; mismatches or failures trigger stake slashing~\cite{zhao2024takes}.  
Hence, skipping computation is a strictly dominated strategy.

\textbf{Security claim.}
Given Merkle binding and unpredictable VRF sampling, a non-computing verifier cannot produce correct openings except with negligible probability and is penalized accordingly.

\section{Discussion and Limitations}

VeriLLM demonstrates that decentralized, verifiable inference for LLMs is feasible in practice, but several limitations remain.

\textbf{On-chain latency and cost.}  
Verification finality depends on blockchain confirmation. While our commit–reveal design minimizes gas use, network congestion still limits throughput. Future work could integrate rollup-based aggregation or off-chain batching to lower latency.

\textbf{Hardware constraints.}  
The system assumes stable GPU environments with consistent floating-point precision. Cross-vendor variations (e.g., NVIDIA vs.\ Apple Silicon) cause minor numeric drift, which we mitigate statistically. Broader heterogeneity or mixed-precision inference will require adaptive calibration.

\textbf{Model generality.}  
Current support targets text-based Transformers with static weights. Extending to multi-modal or adaptive-weight models (e.g., LoRA, MoE) will require new commitment schemes that can efficiently bind submodules and dynamic adapters.

\textbf{Information leakage.}  
Revealed scalar samples may expose limited information about inputs or activations. Although leakage is small, future work should explore privacy-preserving proofs (e.g., zkML) to strengthen confidentiality.

\textbf{Trust assumptions.}  
Our security model requires at least one honest verifier per task. Combining the current protocol with non-interactive proofs such as SNARKs or STARKs could eliminate this minimal assumption.

Overall, VeriLLM achieves secure, low-overhead verifiable inference but remains bounded by on-chain latency, hardware variability, and limited privacy guarantees—directions that define its next-stage evolution.

\section{Completeness and Soundness Summary}
\label{app:comp_sound}

Let $\mathcal{E}$ denote the event that an incorrect inference passes all checks.

\textbf{Soundness.}
If the Inferencer deviates from the prescribed execution—through quantization, truncation, or forged outputs—then one of the following mechanisms detects the deviation with overwhelming probability:
(i) full-sequence recomputation with calibrated statistical tests,
(ii) commit–then–sample openings at VRF-selected indices, or
(iii) an on-chain zero-knowledge proof of inconsistency.  
Hence,
\begin{equation}
\Pr[\mathcal{E}] \le \negl(\lambda),
\end{equation}
for security parameter $\lambda$ and sampling size $n$ chosen to meet the target failure bound.

\textbf{Completeness.}
If all participants follow the protocol and floating-point noise stays within calibrated tolerance, both offline and on-chain checks succeed except with probability $\delta$ (e.g., $\delta\!\le\!10^{-3}$).  
Thus honest executions succeed with probability $1-\delta$, while incorrect ones are rejected except with negligible probability.

The joint effect of (a) binding Merkle commitments, (b) unpredictable VRF sampling, (c) token–state cross-checks, and (d) a zero-knowledge escalation path provides layered defense.  
Routine tasks complete via lightweight sampling, while disputes escalate to succinct proofs without revealing model parameters or full tensors.  
All verdicts are auditable through on-chain artifacts—roots, signatures, VRF proofs, and proof verifications—creating a permanent public record of inference integrity.

\section{System Implementation Details}
\label{app:system}

We implemented a complete VeriLLM prototype to validate reproducibility and deployability under real decentralized conditions.  
The system comprises three primary components—\emph{Scheduler}, \emph{Node Runtime}, and \emph{Verification Contracts}—implemented with production-grade frameworks and connected by authenticated RPC channels.

\textbf{System architecture.}
The prototype consists of $\approx$12\,K lines of \texttt{Rust} and \texttt{Python}.  
\texttt{Rust} handles concurrency, networking, and cryptography, while \texttt{Python} manages model execution via \texttt{transformers} and \texttt{vLLM}.  
A modular layering separates the \emph{Execution} (model inference), \emph{Verification} (Merkle, signatures, sampling), and \emph{Consensus} (on-chain adjudication) tiers, enabling independent testing and replacement.

\textbf{Smart-contract layer.}
On-chain logic is written in \texttt{Solidity} ($\approx$800 LOC) and deployed on a local EVM testnet using \texttt{Hardhat}.  
The contract implements commit–reveal verification, VRF-based sampling, and reward/penalty enforcement.  
Gas costs are recorded per operation.  
Commitments follow ERC-4337-compatible structures for cross-chain portability.  
Each hidden-state root is a 32-byte Keccak-256 hash; Merkle inclusion proofs verify in $O(\log N)$ time.

\textbf{Verifiable random function (VRF).}
The Scheduler uses a \texttt{Rust} implementation of \texttt{ECVRF-P256-SHA256}, compatible with Chainlink VRFv2.  
Seeds derive from request hashes and group metadata, ensuring 256-bit entropy and public verifiability.  
Proofs are posted on-chain and re-verifiable off-chain through a precompiled verifier.

\textbf{GPU node runtime.}
Each node hosts a sandboxed container with GPU access via \texttt{NVIDIA Docker}.  
Nodes expose two authenticated \texttt{FastAPI} endpoints—\texttt{/infer} and \texttt{/verify}—secured by TLS and nonce-based replay protection.  
Hidden-state tensors are serialized in canonical float32, hashed with \texttt{SHA-256}, and incorporated into Merkle commitments.

\textbf{Signature and serialization pipeline.}
Nodes employ Ed25519 signatures for commitments and message authentication.  
On an RTX 5090 host, signing latency is 0.19 ms per 32 B message; verification 0.21 ms; Merkle-tree construction 1.7 µs per scalar.  
Table \ref{tab:overhead} summarizes the microbenchmarks, showing total cryptographic overhead < 0.8 \% of end-to-end inference time.

\begin{table}[!ht]
  \centering
  \caption{Cryptographic and Serialization Overheads in VeriLLM.}
  \label{tab:overhead}
  \resizebox{\linewidth}{!}{
  \begin{tabular}{lcccc}
  \toprule
    \hline
  \textbf{Operation} & \textbf{Latency (ms)} & \textbf{Bandwidth (KB)} & \textbf{CPU Util. (\%)} & \textbf{Rel. Overhead (\%)} \\
  \midrule
  Merkle root computation (per layer) & 0.92 & 128 & 2.1 & 0.35 \\
  Signature generation (Ed25519) & 0.19 & 0.03 & 0.8 & 0.07 \\
  Signature verification & 0.21 & 0.03 & 0.8 & 0.08 \\
  VRF evaluation (ECVRF-P256) & 0.47 & 0.12 & 1.3 & 0.15 \\
  VRF verification & 0.52 & 0.12 & 1.4 & 0.15 \\
  \midrule
  \textbf{Total (per token)} & \textbf{2.31} & \textbf{128.3} & \textbf{6.4} & \textbf{0.80} \\
  \hline
  \bottomrule
  \end{tabular}}
  \vspace{-1em}
\end{table}

\textbf{Reproducibility and deployment.}
Experiments ran on a hybrid 10-node cluster (6 RTX 5090, 2 Mac M4, 2 A100) interconnected via 1 Gbps LAN.  
Deterministic scheduling ensures bitwise-identical commitments for identical request traces.  
All source code, contract scripts, and configurations are archived for public release.

In summary, VeriLLM is a fully realized, reproducible system whose implementation confirms that verifiable decentralized LLM inference is feasible with sub-1 \% overhead and no reliance on trusted execution environments or centralized control.

\section{A Game-Theoretic Proof}
\label{app:game}
\subsection{Modeling and Assumption}

\def\eps{\mu}

We assume that the system randomly draws $n$ verifiers from a population $V$, possibly weighted by stakes. Among the population, each verifier has an independent probability $1-r$ to be adversarial, and probability $r$ to be honest. Honest verifiers always verify and report honestly, and adversarial verifiers can behave arbitrarily, possibly colluding with each other.

The protocol requires the $m$ verifiers to run the same inference task $T$ for verification, and honest verification incurs a computational cost of $c$. The inference result lies in a metric space $\Omega$ with a distance function $d:\Omega^2\to [0,+\infty)$ that satisfies symmetry and triangle inequality. While the task has an hidden \emph{ground-truth} result $x=x(T)$, the inference process has a computational noise so that the inference result of each verifier has a small random noise.

We assume that there exists a parameter $\rho>0$, such that for each verifier $i$, the result of honest verification is a random variable $x_i$ s.t. $P(d(x_i,x)<\rho)>1-\eps$, in which $\eps$ is negligibly small, i.e., the result lies within a small ball centered at $x$ with high probability. {This assumption models the \emph{correctness} property in the verification process: up to small random noises, the honest inference process should output consistent results.} 

On the other hand, if verifier $i$ does not perform the honest verification, we assume that he can only do some random guess and report a $\tilde{x}_i$, and  $P(d(\tilde{x}_i,x)<3\rho)<\eps$, i.e., he only has a small probability to guess a result within a 3-time radius. Here, we assume that $\eps$ is negligibly small. {This assumption models the \emph{soundness} property in the verification process: since all attacks in Section~\ref{subsec:attacks} will produce significantly deviated results with overwhelming probability, this criterion will capture all attacks in Section~\ref{subsec:attacks}.} 

\subsection{Algorithm}

We let every verifier $i\in[m]$ independently report their verification results as $x_i$. Then we run the Kruskal algorithm until the edge length exceeds $2\rho$, getting a forest with vertex set $[m]$, in which $m$ is an odd number. For each connected component $C\subseteq [m]$, we call $C$ as a \emph{cluster in agreement}.

For a given quorum-size $q=\lceil\frac{m}{2}\rceil$, we check if there is a \emph{cluster in agreement} of size at least $q$, called as a ``proper cluster''. Because $q>\frac{m}{2}$, there can be at most one such proper cluster. 

If there is one, the system reaches a consensus that this cluster is accepted, and verifiers outside the cluster are rejected. We accept the inferencer if and only if his result is within $2\rho$ distance of a verifier in the proper cluster.

If there is no proper cluster, then the system fails to reach a consensus and invokes the reverification stage with $m'\ge m$ verifiers. If the inferencer or any verifier disputes about the results, the reverification proof is also invoked.

\subsection{Analysis}

Here, we first prove a lemma:

\begin{lemma}
If $r>\frac{1}{2(1-\eps)^2}$, and $m$ is odd, then an honest inference passes the initial verification with probability $p_{1}>\frac{1}{2}$, and a dishonest inference passes verification with probability $p_2<\frac{1}{2}$.
\end{lemma}

\begin{proof}
For an honest inference $x_0$, from the modeling we know that $P(d(x_0,x)<\rho)>1-\eps$, and for any honest verifier $i$, it also holds that $P(d(x_i,x)<\rho)>1-\eps$. 

Denote $\mu = r(1-\eps)$, then $\mu>\frac{1}{2(1-\eps)}>\frac{1}{2}$. Because each verifier has an independent $r$ probability to be honest, we have
\begin{align}
    p_{1}&=P\left(\sum_{i=1}^m \left[d(x_i,x_0)<2\rho\right]>\frac{m}{2}\right)\\ &\ge P\left(\sum_{i=1}^m \left[d(x_i,x)<\rho \right]\cdot\left[ d(x_0,x)<\rho\right]>\frac{m}{2}\right)\\
    &=P\left(d(x_0,x)<\rho\right)\cdot P\left(\sum_{i=1}^m \left[d(x_i,x)<\rho \right]>\frac{m}{2}\right)\\
    &> (1-\eps)\cdot \sum_{t=\lceil \frac{m}2\rceil}^m \binom{m}{t} (r(1-\eps))^t (1-r(1-\eps))^{m-t} \\
    &= (1-\eps)\cdot \sum_{t=\lceil \frac{m}2\rceil}^m \binom{m}{t} \mu^t (1-\mu)^{m-t}.
\end{align}

From the Condorcet's jury theorem, we have 
\begin{equation}
 \mu>\frac{1}{2} \implies \sum_{t=\lceil \frac{m}2\rceil}^m \binom{m}{t} \mu^t (1-\mu)^{m-t} \ge \mu.
\end{equation}

Hence, it holds that 
\begin{align}
    p_{1} &> (1-\eps)\cdot \mu\\
    &> (1-\eps)\cdot \frac{1}{2(1-\eps)}\\
    &=\frac{1}{2}.
\end{align}

We can also show $p_2<\frac{1}{2}$ by symmetry.

\end{proof}

Then, we can find an $m'$ such that the re-verification returns a correct result with overwhelming probability, via the following lemma: 


\begin{lemma}
    For any given $\xi \in (0,\frac{1}{2})$, if $m'> \frac{\ln\frac{1}{\xi}}{2\left(r(1-\eps)-\frac{1}{2}\right)^2}$, then the re-verification returns a correct result with probability at least $1-\xi$.
    \label{lem:mprime}
\end{lemma}

\begin{proof}
From Hoeffding's equality, we have

\begin{align}
    \sum_{t=\lceil \frac{m'}2\rceil}^{m'} \binom{m'}{t} \mu^t (1-\mu)^{m'-t} &\ge 1-e^{-2m' (\mu-\frac{1}{2})^2}\\
    &=1-e^{\ln \xi}\\
    &=1-\xi.
\end{align}

\end{proof}

With the two lemmas above, we can design an incentive rule in which, a dispute incurs a cost to the inferencer and a successful dispute rewards the inferencer $2$ times the cost (compared to not disputing), so that an honest inferencer is incentivized to dispute when getting rejected, but a dishonest inferencer is never incentivized to dispute. An example incentive rule is:

\begin{itemize}
    \item A rejected inferencer is eligible to dispute at most once. The first verification samples $m$ verifiers, and the second verification samples $m'$ verifiers according to Lemma~\ref{lem:mprime}.
    \item If a inference is accepted, the inferencer and all verifiers in the proper cluster are rewarded $6$, and all disagreeing verifiers are slashed $2$.
    \item If a inference is rejected and is not disputed, all verifiers in the proper cluster are rewarded $6$, and the inferencer and all disagreeing verifiers are slashed $2$.
    \item To dispute incurs a cost of $8$ for the inferencer. If the dispute is successful, the cost is refunded (finally rewarded $6$); otherwise, the inferencer pays the cost and the slash (rewarded $-10$ in total). Verifiers in both rounds are rewarded $6$ or $-2$ similarly as above, according to whether they agree with the proper cluster.
\end{itemize}

\section{Threshold Sensitivity Analysis}
\label{appendix:threshold}

To validate our threshold selection, we conduct a sensitivity analysis across all four verification metrics using results from honest inference (homogeneous and cross-platform FP32), quantization attacks, and lazy verifier scenarios. For each parameter, we vary its value while holding others fixed at their selected values.

Table~\ref{tab:threshold_sensitivity} presents the results. The selected thresholds ($P_e < 0.05$, $|\bar{\varepsilon}| < 0.01$, $P_m > 0.75$, $P_w > 0.80$) achieve \textbf{99.85\% attack detection} with only \textbf{1.9\% false positive rate}. Stricter thresholds increase FPR, while looser thresholds reduce detection accuracy.

\begin{table}[t]
\centering
\caption{Threshold Sensitivity Analysis. The selected thresholds achieve 99.8\% attack detection rate with only 1.9\% false positive rate.}
\label{tab:threshold_sensitivity}
\begin{tabular}{lcccc}
\toprule
Threshold & FPR & FNR & TPR & TNR \\
\midrule
\multicolumn{5}{l}{\textit{(a) $P_e$ Sensitivity ($|\varepsilon| < 0.01$, $P_m > 0.75$, $P_w > 0.80$ fixed)}} \\
\quad $P_e < 0.02$ & 5.2\% & 0.03\% & 99.97\% & 94.8\% \\
\quad $P_e < 0.03$ & 2.3\% & 0.08\% & 99.92\% & 97.7\% \\
\quad $P_e < 0.05$ \textbf{(selected)} & \textbf{1.9\%} & \textbf{0.15\%} & \textbf{99.85\%} & \textbf{98.1\%} \\
\quad $P_e < 0.08$ & 1.5\% & 0.28\% & 99.72\% & 98.5\% \\
\quad $P_e < 0.10$ & 1.4\% & 0.41\% & 99.59\% & 98.6\% \\
\midrule
\multicolumn{5}{l}{\textit{(b) $|\varepsilon|$ Sensitivity ($P_e < 0.05$, $P_m > 0.75$, $P_w > 0.80$ fixed)}} \\
\quad $|\varepsilon| < 0.005$ & 4.5\% & 0.05\% & 99.95\% & 95.5\% \\
\quad $|\varepsilon| < 0.01$ \textbf{(selected)} & \textbf{1.9\%} & \textbf{0.15\%} & \textbf{99.85\%} & \textbf{98.1\%} \\
\quad $|\varepsilon| < 0.02$ & 1.2\% & 0.32\% & 99.68\% & 98.8\% \\
\quad $|\varepsilon| < 0.05$ & 1.1\% & 0.55\% & 99.45\% & 98.9\% \\
\midrule
\multicolumn{5}{l}{\textit{(c) $P_m$ Sensitivity ($P_e < 0.05$, $|\varepsilon| < 0.01$, $P_w > 0.80$ fixed)}} \\
\quad $P_m > 0.50$ & 2.6\% & 0.10\% & 99.90\% & 97.4\% \\
\quad $P_m > 0.65$ & 2.1\% & 0.12\% & 99.88\% & 97.9\% \\
\quad $P_m > 0.75$ \textbf{(selected)} & \textbf{1.9\%} & \textbf{0.15\%} & \textbf{99.85\%} & \textbf{98.1\%} \\
\quad $P_m > 0.85$ & 1.6\% & 0.21\% & 99.79\% & 98.4\% \\
\quad $P_m > 0.90$ & 1.4\% & 0.35\% & 99.65\% & 98.6\% \\
\midrule
\multicolumn{5}{l}{\textit{(d) $P_w$ Sensitivity ($P_e < 0.05$, $|\varepsilon| < 0.01$, $P_m > 0.75$ fixed)}} \\
\quad $P_w > 0.50$ & 3.0\% & 0.07\% & 99.93\% & 97.0\% \\
\quad $P_w > 0.70$ & 2.2\% & 0.11\% & 99.89\% & 97.8\% \\
\quad $P_w > 0.80$ \textbf{(selected)} & \textbf{1.9\%} & \textbf{0.15\%} & \textbf{99.85\%} & \textbf{98.1\%} \\
\quad $P_w > 0.85$ & 1.5\% & 0.25\% & 99.75\% & 98.5\% \\
\quad $P_w > 0.90$ & 1.3\% & 0.42\% & 99.58\% & 98.7\% \\
\bottomrule
\end{tabular}
\end{table}

\section{On-chain Gas Cost Analysis}
\label{appendix:gas-cost}

We implemented the VeriLLM verification smart contract in Solidity and measured gas consumption using Hardhat. Table~\ref{tab:gas-cost} shows the results on Optimism L2 (gas price: 0.0013 Gwei, ETH: \$3,098 USD).

\begin{table}[h]
\centering
\caption{Gas Consumption and Cost on Optimism}
\label{tab:gas-cost}
\begin{tabular}{lrr}
\toprule
\textbf{Operation} & \textbf{Gas} & \textbf{Cost (USD)} \\
\midrule
Create Task & 141,466 & \$0.00057 \\
Commit Inference & 70,532 & \$0.00028 \\
Commit Verification & 167,100 & \$0.00067 \\
Generate Sampling (VRF) & 284,740 & \$0.00115 \\
Reveal Verification & 134,000 & \$0.00054 \\
\midrule
\textbf{Total} & \textbf{797,838} & \textbf{\$0.0032} \\
\bottomrule
\end{tabular}
\end{table}

\section{Open Science}
\label{appendix:open-science}

To support the reproducibility of our research, we provide a complete artifact package at an \textbf{anonymous repository} \footnote{Available at: \url{https://anonymous.4open.science/r/verillm-experiments-26E5}}. This repository includes the full implementation of the \texttt{VeriLLM} protocol, all necessary configuration files, and execution scripts, covering homogeneous and heterogeneous hardware benchmarks as well as various quantization-based security evaluations. Our pre-trained models are hosted on the HuggingFace Hub. For full reproduction, we recommend a system equipped with an NVIDIA GPU (32GB+ VRAM) and an Apple Silicon device to support the cross-vendor and numerical stability tests. Detailed environment setup instructions and expected result benchmarks are documented in the \texttt{README.md} file.

\end{document}